\documentclass[journal]{IEEEtran}

\usepackage{xcolor,soul,framed} %,caption

\colorlet{shadecolor}{yellow}
\usepackage[pdftex]{graphicx}
\graphicspath{{../pdf/}{../jpeg/}}
\DeclareGraphicsExtensions{.pdf,.jpeg,.png}

\usepackage[cmex10]{amsmath}
\usepackage{array}
\usepackage{mdwmath}
\usepackage{mdwtab}
\usepackage{eqparbox}
\usepackage{url}

\usepackage{multirow}
\usepackage{multicol}

\let\labelindent\relax
\usepackage{enumitem}
\usepackage{multirow}
\usepackage{multicol}
\usepackage{enumitem}

\usepackage{numprint}
\npthousandsep{,}

\usepackage{graphicx}
\usepackage{array, makecell}
\usepackage{verbatim}
\usepackage[utf8]{inputenc}
\usepackage[T1]{fontenc}
\usepackage[ngerman,english]{babel}
\usepackage{tabularx}  % for 'tabularx' environment and 'X' column type
\usepackage{ragged2e}  % for '\RaggedRight' macro (allows hyphenation)
\newcolumntype{Y}{>{\RaggedRight\arraybackslash}X} 
\usepackage{booktabs}  % for \toprule, \midrule, and \bottomrule macros 

\usepackage{tabularray}

\usepackage{xcolor}
\usepackage[normalem]{ulem}
\usepackage{circledsteps}
\usepackage{svg}
\usepackage{caption}
\usepackage{subcaption}
\usepackage{tabularray}

\definecolor{additioncolor}{RGB}{32,138,0}
\definecolor{deletioncolor}{RGB}{179,0,0}

\def\showrevisions{1}

\ifx\showrevisions\undefined
    
    \newcommand{\deletion}[1]{}
    
\else
    
    \newcommand{\deletion}[1]{\textcolor{deletioncolor}{\ifmmode\text{\sout{\ensuremath{#1}}}\else\sout{#1}\fi}}
    
\fi

\newcommand{\pitchcorr}{\boldsymbol{\rho}^{F_0}}
\newcommand{\gvd}{G_{\text{VD}}}
\newcommand{\utt}{\mathbf{u}}
\newcommand{\newpart}[1]{#1}
\newcommand{\newparttwo}[1]{#1}

\newcommand{\figref}[1]{Figure~\ref{#1}}
\newcommand{\secref}[1]{Section~\ref{#1}}

\begin{document}
\bstctlcite{IEEEexample:BSTcontrol}
    \title{The VoicePrivacy 2022 Challenge: Progress and Perspectives in Voice Anonymisation}
    \author{
    Michele~Panariello,$^*$~\IEEEmembership{Student Member,~IEEE,}
    Natalia~Tomashenko,$^*$~\IEEEmembership{Member,~IEEE,}
    Xin~Wang,$^*$~\IEEEmembership{Member,~IEEE,}
    Xiaoxiao~Miao,~\IEEEmembership{Member,~IEEE,}
    Pierre~Champion,
    Hubert~Nourtel,
    Massimiliano~Todisco,~\IEEEmembership{Member,~IEEE,}
    Nicholas~Evans,~\IEEEmembership{Member,~IEEE,}
    Emmanuel~Vincent,~\IEEEmembership{Fellow,~IEEE,}
    Junichi~Yamagishi,~\IEEEmembership{Senior~Member,~IEEE}%
  % \author{x~y,~\IEEEmembership{Member,~IEEE,}
      % a~b,~\IEEEmembership{Member,~IEEE,}\\
      % Ignacio~Ramos,~\IEEEmembership{Student Member,~IEEE,}
      % Erez Falkenstein,~\IEEEmembership{Student Member,~IEEE,}
      % and~Zoya~Popovi\'c,~\IEEEmembership{Fellow,~IEEE}% <-this % stops a space
}  

% alphabetical list from the website:
% Jean-François Bonastre                prefers not to be included - email 15/12

% Pierre Champion                          confirmed. ORCID: 0000-0003-1225-7957
% Nicholas Evans, Member, IEEE             confirmed. ORCID: 0000-0002-8459-1041
% Xiaoxiao Miao, Member, IEEE              confirmed. ORCID: 0000-0002-6645-6524
% Hubert Nourtel                           confirmed. ORCID: 0009-0001-8760-0046
% Massimiliano Todisco, Member, IEEE       confirmed. ORCID: 0000-0003-2883-0324
% Natalia Tomashenko, Member, IEEE         confirmed. ORCID: 0000-0002-7125-2382
% Emmanuel Vincent, Senior Member, IEEE    confirmed. ORCID: 0000-0002-0183-7289
% Xin Wang, Member, IEEE                   confirmed. ORCID: 0000-0001-8246-0606
% Junichi Yamagishi, Senior Member, IEEE   confirmed. ORCID: 0000-0003-2752-3955

% ---

% Michele Panariello ORCID: 0009-0007-4154-5460

\markboth{IEEE/ACM TRANSACTIONS ON AUDIO, SPEECH, AND LANGUAGE PROCESSING, VOL.~X, NO.~X, MONTH~YEAR
}{Author \MakeLowercase{\textit{et al.}}: The VoicePrivacy 2022 Challenge: Progress and Perspectives in Voice Anonymisation}

% ====================================================================
\maketitle

\def\thefootnote{*}\footnotetext{The first three authors contributed equally to this work.}\def\thefootnote{\arabic{footnote}}

% === ABSTRACT ====================================================================
% =================================================================================
\begin{abstract}
The VoicePrivacy Challenge promotes the development of voice anonymisation solutions for speech technology. 
In this paper we present a systematic overview and analysis of the second edition held in 2022.
We describe the voice anonymisation task and datasets used for system development and evaluation, present the different attack models used for evaluation, and the associated objective and subjective metrics.
We describe three anonymisation baselines, provide a summary description of the anonymisation systems developed by challenge participants, and report objective and subjective evaluation results for all.
In addition, we describe post-evaluation analyses and a summary of related work reported in the open literature.
% Given mounting privacy regulation worldwide, we expect voice anonymisation, in addition to privacy preservation more generally, to attract far greater attention in the coming years.
Results show that solutions based on voice conversion better preserve utility, that an alternative which combines automatic speech recognition with synthesis achieves greater privacy, and that a privacy-utility trade-off remains inherent to current anonymisation solutions.
%; there is currently no silver bullet which delivers both. 
Finally, we present our ideas and priorities for future VoicePrivacy Challenge editions.
\end{abstract}

% === KEYWORDS ====================================================================
% =================================================================================
\begin{IEEEkeywords}
anonymisation, pseudonymisation, voice privacy, speech synthesis, voice conversion, attack model %speaker verification, speech recognition
\end{IEEEkeywords}

% \fi
\IEEEpeerreviewmaketitle

\section{Introduction}
%\IEEEPARstart{A}{nonymization} is one approach to preserve privacy when speech signals are shared with others~\cite{Tomashenko2021CSl}.  
%The need for privacy preservation has come to the fore in recent years due to the rise of smart voice interactive technology and services, and also in response to mounting privacy regulation.  

\IEEEPARstart{S}{peech} data contain extensive personal, sensitive information which goes far beyond the spoken message. 
The speaker identity, health and emotional condition, socio-economic status, geographical origin, among a host of other attributes, can all be estimated from speech recordings~\cite{Nautsch-PreservingPrivacySpeech-CSL-2019, tomashenko2020introducing}.
Without safeguards, all such information is potentially disclosed as soon as speech signals are shared.
Even when consent is given to the use of speech data for a specific voice service, e.g.~those provided by a smart speaker, there is no guarantee that it will not also be used for other purposes. 

Privacy can be preserved by sanitising a speech recording of specific personal, sensitive information before it is shared.
While the community has made inroads in recent years to develop approaches to disentangle and suppress different sources of such information, effective and comprehensive solutions have yet to be developed. 
One branch of research in privacy preservation in which progress 
has been rapid in recent years involves \emph{anonymisation}, namely the suppression of personally identifiable information (PII) or cues which can be used by human listeners and/or automatic systems to infer identity.

PII includes information carried by the pitch or fundamental frequency, the timber or spectral envelope, distinctive spoken content (e.g.~the speaker's name or social security number),  para-linguistic traits and background sounds, among other sources.
While a comprehensive approach calls for \emph{all} such sources of personal, sensitive information (or a selected subset) to be suppressed or masked, thus far the community has focused predominantly upon \emph{voice anonymisation}\footnote{Note that, in the legal community, the term ``anonymisation'' means that this goal has been achieved. Here, it refers to the task to be addressed, even when the method being evaluated has failed.}~\cite{Tomashenko2021CSl}.  %Voice anonymisation 
It refers to the substitution of a speaker's own, natural voice with that of another, pseudo-speaker, while leaving all other attributes (e.g.\ linguistic content and para-linguistic attributes) intact~\cite{Tomashenko2021CSl}.
Privacy is assumed to be preserved if any remaining sensitive content can no longer be linked to the original speaker through the use of traditional voice-related cues.
\newparttwo{Note that voice anonymisation focuses exclusively on concealing the \emph{acoustic} characteristics of the speaker voice and is not concerned with other sources of PII which can also be inferred (e.g.~the spoken content of the speech).}

The first VoicePrivacy Challenge held in 2020 showed that, while voice anonymisation can improve privacy, the technology at the time was far away from the goal of delivering full anonymisation.
Results showed that the voice identity can still be revealed, albeit with greater difficulty, under a semi-informed attack model, and that more robust anonymisation cannot be achieved without degrading utility.
%to u e constraint of utility preservation makes it challenging to design more effective defense mechanisms.
The second challenge edition was organised in 2022, with a view to bolstering research effort in the field and to fostering progress in the development of more reliable, effective voice anonymisation solutions.
The VoicePrivacy 2022 workshop was held in conjunction with the 2nd Symposium on Security and Privacy in Speech Communication (SPSC),\footnote{\url{https://symposium2022.spsc-sig.org/}} a joint event co-located with Interspeech~2022.
%Voice anonymisation and voice privacy, in addition to broader interests in privacy preservation for speech technology, are now gaining considerable traction.
Reported in this paper are the principal findings from the challenge results and discussions during the workshop. 
We describe the privacy preservation scenario addressed with VoicePrivacy, the evaluation methodology, 
%an as-yet-unpublished 
a summary of the baselines, competing systems and results, together with a treatment of post-evaluation work and a roadmap for future research directions and priorities.
The challenge results show that no system excelled in all evaluation metrics and that there is currently no silver bullet solution to voice anonymisation. 
%Moreover, objective and subjective metrics often do not align. We conclude that the evaluation protocol should be revisited to target practical use cases, encouraging the convergence of research on speaker anonymisation and real-world applications.

\section{Challenge design}
%In this section, 
We present 
%an overview of the  challenge : 
the scenario and requirements, the anonymisation task, and the attack models, protocol and datasets.
% , data and 
% evaluation methodology. 
% We present a brief overview of the VoicePrivacy~2022 Challenge and design.  We describe: the scenario and requirements; the anonymisation task; the attack models; the protocol and datasets.
%; the evaluation methodology. 

%\subsection{anonymisation task and attack models}
\subsection{Scenario and requirements}
\label{sec:task}

%\subsubsection{Scenario and task}\label{subsec:anon_task}
The scenario includes two actors -- a user who wishes to preserve privacy using an anonymisation safeguard, and an adversary who wishes to undermine the same safeguard.
The user wishes to post online or otherwise share an audio recording. 
It contains speech in his or her voice (or perhaps that of some other individual\footnote{
Each recording is assumed to contain the speech of a single individual.  
The anonymisation of recordings containing multiple voices can be also be achieved, e.g.~using speaker diarization and the separate application of anonymisation to the set of segments corresponding to each speaker/voice.}) 
and is shared in order to accomplish some downstream task, e.g.~to share personal content via social media or to access a voice interactive information service.
%, or to contribute to speech data collection initiatives such as for training of automatic speech recognition systems.
The individual wishes to preserve privacy and guard against the misuse of recordings by fraudsters for malicious purposes, e.g.~the training of voice conversion or speech synthesis models which would allow the generation of artificial speech recordings (spoofs/deepfakes) in their voice.  

Before sharing, the recordings are anonymised so as to ensure, to the extent possible, that they cannot be linked to the speaker whose voice they contain.  
%The utterances shared by the users are referred to as \emph{trial} utterances. 
%In order to hide his/her identity, each user passes these utterances through a voice anonymisation system prior to sharing. 
The speech in an anonymised recording should hence contain the voice of a different individual referred to as a \emph{pseudo-speaker}.  
The pseudo-speaker might, for instance, have an artificial voice which does not correspond to any real speaker.
Nonetheless, the voice of the same pseudo-speaker should always be used for recordings of the same, original speaker. 
This requirement stems from the eventual use of anonymisation in multi-speaker scenarios in which the voice of different individuals should remain distinctive.

While the voice identity should be masked, other speech characteristics should be preserved.  
These include both linguistic and para-linguistic attributes.
%such as the intonation.  
The preservation of linguistic content -- the spoken words -- is paramount to almost any conceivable downstream task and hence intelligibility should be preserved.  
The requirement to preserve other, para-linguistic attributes is more dependent upon the specific downstream task (e.g.\ automatic speech recognition, speaker recognition or diarization, emotion analysis, etc.).
To promote the development of anonymisation solutions for reasonably diverse downstream tasks, prosody (intonation, stress, rhythm etc.) should also be preserved.

The privacy adversary seeks to undermine the safeguard and to \emph{re-identify} the original speaker whose voice is contained in an anonymised recording. 
%The attacker is assumed to have access to one or more recordings which contain the genuine voice of an individual whose voice is suspected to correspond to the anonymised recording.
To do so, the attacker makes comparisons between recordings which contain genuine, unprotected voices and recordings containing anonymised voices. 
An effective anonymisation system should make re-identification -- the linking of anonymised, pseudo-voices to genuine, unprotected voices -- as difficult as possible.
\newparttwo{Note that the potential for an attacker to re-identify the speaker is gauged based solely on the use of acoustic voice characteristics, and does not take into account any additional sources of PII such as that contained in the spoken content.}

\subsection{Anonymisation task} \label{sub:anon_task}

The task of VoicePrivacy challenge participants is to develop anonymisation systems which fulfil the requirements outlined above. They should: 
\begin{enumerate}[label=(\alph*)]
    \item output a speech waveform; 
    \item conceal the speaker identity; 
    \item preserve linguistic and para-linguistic attributes; 
    \item ensure that all utterances corresponding to one speaker are anonymised so that they contain the voice of the same pseudo-speaker, while utterances corresponding to different speakers are anonymised so that they contain the voice of different pseudo-speakers.
\end{enumerate}

We define the latter condition as \emph{speaker-level} anonymisation.
This is the default requirement for VoicePrivacy challenges.
It is different to \emph{utterance-level} anonymisation for which each utterance is anonymised using \emph{different} pseudo-voices, even when they are produced by the same speaker.

\subsection{Attack models}
\label{subsec:attack_model}

The resources and information that are available to the privacy adversary, and the efforts to which the adversary goes in order to undermine the anonymisation safeguard are defined in the form of an \emph{attack model}.
The adversary is assumed to be anonymisation-aware and will adapt in order to increase the chances of re-identifying the speaker.
The adversary is furthermore assumed to have access to one or more recordings which contain the genuine, unprotected voice of an individual whose voice is suspected to correspond to the anonymised recording. 
The adversary can hence make comparisons between the unprotected recordings and the anonymised/protected recordings to infer identity.

The adversary may decide to anonymise the unprotected recordings to reduce domain mismatch during the comparison.
From hereon, and in the vein and terminology of automatic speaker verification (ASV), recordings that are anonymised by users who wish to protect their identity are referred to as \emph{trial utterances}, whereas recordings used by privacy adversaries to undermine or reverse the anonymisation are referred to as \emph{enrolment utterances}.
The attacker uses an ASV system to verify whether or not the voices in trial and enrolment utterances correspond to that of the same individual.

The attacker is assumed to have access to the same anonymisation system as the user,
but not their specific configuration or system parameters. 
This is a reasonable assumption when the anonymisation system is available to many different users and embraces a worse-case scenario for assessment, i.e.\ a \emph{strong} attack model.
To improve the potential of re-identifying (or not) the original speaker of a trial utterance, the attacker can use the system to anonymise the enrolment utterance in order to reduce the domain mismatch between it and the anonymised trial utterance. 
This is achieved using a large set of similarly anonymised data which the adversary can use to train a new ASV system, or adapt an existing system, optimised to operate upon anonymised data.

\begin{table}[!t]
\centering
  \caption{Number of speakers and utterances in the training, development, and evaluation sets~\cite{tomashenko2020introducing}.}\label{tab:data}
 \resizebox{0.49\textwidth}{!}{
  \centering
  \begin{tabular}{|c|l|l|r|r|r|r|}
\hline
 \multicolumn{3}{|l|}{\textbf{Subset}} &  \textbf{Female} & \textbf{Male} & \textbf{Total} & \textbf{\#Utterances}  \\ \hline \hline
% train
\multirow{5}{*}{\rotatebox{90}{Training~}} & \multicolumn{2}{l|}{VoxCeleb-1,2} & \numprint{2912} & \numprint{4451} & \numprint{7363} & \numprint{1281762} \\ \cline{2-7}
& \multicolumn{2}{l|}{LibriSpeech train-clean-100} & 125 & 126 & 251	& \numprint{28539} \\ \cline{2-7}
& \multicolumn{2}{l|}{LibriSpeech train-other-500} & 564 & 602 & \numprint{1166} & \numprint{148688}	\\ \cline{2-7}
& \multicolumn{2}{l|}{LibriTTS train-clean-100} & 123 & 124 & 247 & \numprint{33236} \\ \cline{2-7}
& \multicolumn{2}{l|}{LibriTTS train-other-500} & 560 & 600 & \numprint{1160} & \numprint{205044} \\ \hline\hline
% devel
\multirow{5}{*}{\rotatebox{90}{~~~Dev.}} & LibriSpeech & Enrollment & 15 & 14 & 29 & 343\\ \cline{3-7}
& dev-clean & Trial & 20 & 20 & 40 & \numprint{1978}\\ \cline{2-7}
& \multirow{2}{*}{VCTK-dev} & Enrollment & \multirow{2}{*}{15} & \multirow{2}{*}{15} & \multirow{2}{*}{30} & 600 \\  \cline{3-3}\cline{7-7}
% & VCTK-dev & Trial (common) & 15 & 15 & 30 & 695\\  \cline{3-3}\cline{7-7}
% & & Trial (different) & & & & \numprint{10677} \\  \hline\hline
& & Trial & & & & \numprint{11372} \\  \cline{3-3}\cline{7-7} % the value here is 10677 (different) + 695 (common)
% & & Trial (common) & & & &  695\\
\hline\hline
% eval
\multirow{5}{*}{\rotatebox{90}{~~~Eval.}} & LibriSpeech & Enrollment & 16 & 13 & 29 & 438\\ \cline{3-7}
& test-clean & Trial & 20 & 20 & 40 & \numprint{1496}\\ \cline{2-7}
& \multirow{2}{*}{VCTK-test} & Enrollment & \multirow{2}{*}{15} & \multirow{2}{*}{15} & \multirow{2}{*}{30} & 600 \\ \cline{3-3}\cline{7-7}
% & VCTK-test & Trial (common) & 15 & 15 & 30 & 70 \\ \cline{3-3}\cline{7-7}
% & & Trial (different) & & & & \numprint{10748} \\ \hline
& & Trial & & & & \numprint{11448} \\ \cline{3-3}\cline{7-7} % the total here is 10748 (vtctk different) + 700 (vctk common)
% & & Trial (common) & & & & 700 \\
\hline
\end{tabular}}
\end{table}

\subsection{Protocols and datasets}\label{sec:data}
A set of protocols was designed and made available to VoicePrivacy 2022 participants so that solutions developed using the same data resources can be meaningfully compared. 
A set of publicly-available datasets are used for training, development and evaluation and are the same as those used for the inaugural 2020 challenge edition~\cite{Tomashenko2021CSl}.  
They comprise the distinct, non-overlapping sets presented in Table~\ref{tab:data}, all of which are composed of multiple corpora as follows.

\paragraph{Training set}
The training set comprises the \numprint{2800}~h \textit{VoxCeleb-1,2} speaker verification corpora~\cite{nagrani2017voxceleb,chung2018voxceleb2} and 600~h subsets of the \textit{LibriSpeech} \cite{panayotov2015librispeech} and \textit{LibriTTS} \cite{zen2019libritts} corpora.
See Table~\ref{tab:data} for more details.
Challenge participants were permitted to use this data only for the training of an anonymisation system. %The usage of further training data was not allowed.

\paragraph{Development set}
The development set comprises the \textit{LibriSpeech dev-clean} dataset and a subset of the \textit{VCTK} dataset~\cite{yamagishi2019cstr} denoted~\textit{VCTK-dev}.
Both are split into trial and enrolment subsets which are used in the manner described in Section~\ref{subsec:attack_model} above.
Two development datasets are used in order to test anonymisation performance under matched and mis-matched conditions; the training partition contains data sourced from the \textit{LibriSpeech} dataset only.
%\footnote{The degree of overlap between enrolment and trial speakers is different for \textit{LibriSpeech} and \textit{VTCK} datasets.} % but
%, since we observed no notable, corresponding differences in anonymisation performance,
%it is not described here.}
% While there are differences in the speaker and utterance overlap between the \textit{LibriSpeech dev-clean} and \textit{VCTK-dev} databases, they are not described here; we observed no notable, corresponding differences in anonymisation performance.  
Full details can be found in the VoicePrivacy 2022 evaluation plan~\cite{tomashenko2022voiceprivacy}.
Details of the development protocol
are shown in the middle of Table~\ref{tab:data}.
%VCTK subset is \emph{out-of-domain} in the sense that  development partition, since no subset of VCTK is included in the training data.
%For the \textit{LibriSpeech dev-clean} dataset, speakers in the enrolment subsets are a subset of those in the trial set. 
%For \textit{VCTK-dev}, speakers in the enrolment and trial subsets are the same.
%The \textit{VCTK-dev} data is further split into two trial subsets: \textit{common} (identical utterances for all speakers); \textit{different} (distinct utterances for all speakers).
%The enrollment and \textit{different} subsets comprise distinct utterances for all speakers.

\paragraph{Evaluation set} 
The evaluation set comprises the \textit{LibriSpeech test-clean} dataset and another subset of the \textit{VCTK} dataset denoted \textit{VCTK-test}.
% should we say again that the two subsets are there to test the system under different conditions? I guess not...
Details are shown to the bottom of Table~\ref{tab:data}.
\newpart{Both development and evaluation sets contain exclusively utterances in English. The majority of the training set is also in English, however VoxCeleb1-2 also contain speech from non-English speakers~\cite{nagrani2017voxceleb, chung2018voxceleb2}.}

%Challenge participants use development and evaluation sets in the usual way.
%Only trial utterances are anonymised; enrollment utterances are used to assess anonymisation performance.

\section{Metrics}

A set of objective and subjective \emph{privacy} and \emph{utility} metrics were adopted for the VoicePrivacy challenge series to assess voice anonymisation performance
%Objective and subjective \emph{utility} metrics are used to assess 
and the fulfilment of user goals or downstream tasks respectively. Example downstream tasks are described in Section~\ref{sec:task}. 
A suite of evaluation tools and scripts is freely available.\footnote{\label{fn:git}Evaluation scripts: \url{https://github.com/Voice-Privacy-Challenge/Voice-Privacy-Challenge-2022}}

\subsection{Primary objective assessment}
\subsubsection{Privacy -- equal error rate (EER)}\label{sec:asv-eval}
% \begin{figure}[t!]
% \centering\includegraphics[width=\columnwidth]{Figures/ASV_ep_no_line.pdf} % original width was 89mm
% \caption{ASV evaluation: an \textit{unprotected} scenario (top) in which an entirely conventional $ASV_\text{eval}$ system trained on unprotected data operates upon unprotected trial and enrollment utterances; 
% an \textit{semi-informed} attack scenario (bottom) in which an $ASV_\text{eval}^{\text{anon}}$ system trained on \textit{utterance-level} anonymised data operates upon \textit{speaker-level} anonymised trial and enrollment data.}
% %with different pseudo-speakers,  }
% \label{fig:asv-eval}
% \end{figure}

\begin{figure}[t!]
\centering\includegraphics[width=\columnwidth]{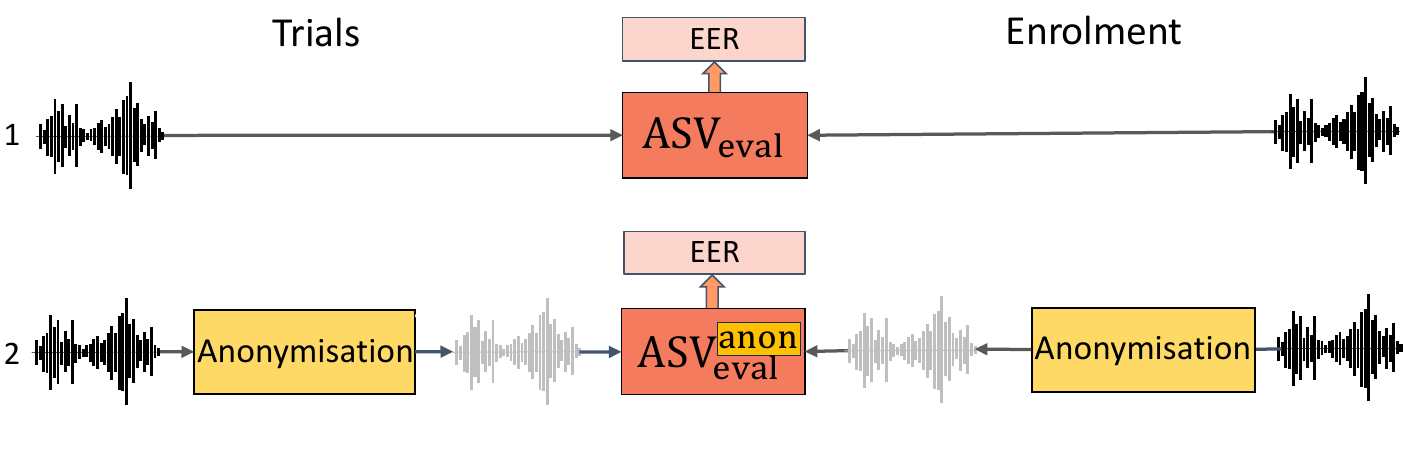} % original width was 89mm
\caption{ASV evaluation (1) \textit{unprotected}: original trial and enrollment data, $ASV_\text{eval}$ trained on original data;
(2)~\textit{semi-informed} attacker: \textit{speaker-level} anonymised trial and enrollment data with different pseudo-speakers,  $ASV_\text{eval}^{\text{anon}}$ trained on \textit{utterance-level} anonymised data.}
%with different pseudo-speakers,  }
\label{fig:asv-eval}
\end{figure}

\begin{table}[t]
  \caption{Number of speaker verification trials.}\label{tab:trials}
  \renewcommand{\tabcolsep}{0.13cm}
  \centering
   \resizebox{0.49\textwidth}{!}{
  \begin{tabular}{|l|l|l|r|r|r|}
\hline
 \multicolumn{2}{|l|}{\textbf{Subset}} & \textbf{Trials} &  \textbf{Female} & \textbf{Male} & \textbf{Total}  \\ \hline \hline
% dev
\multirow{6}{*}{\rotatebox{90}{~~~~Dev.}} & LibriSpeech & Same-speaker & 704 & 644 & \numprint{1348} \\ \cline{3-6}
 & dev-clean & Different-speaker	& \numprint{14566} & \numprint{12796} &	\numprint{27362} \\ \cline{2-6}
& \multirow{2}{*}{VCTK-dev} & Same-speaker & \numprint{2125} & \numprint{2366} & \numprint{4491} \\ \cline{3-6}
 % & & Same-speaker (common) & \numprint{344} & \numprint{351} & \numprint{695} \\  \cline{3-6}
 & & Different-speaker & \numprint{18029} & \numprint{17896} & \numprint{35925} \\ % \cline{3-6}
 % & & Different-speaker (common) & \numprint{4810} &	\numprint{4911} & \numprint{9721} \\
 \hline\hline
% eval
\multirow{6}{*}{\rotatebox{90}{~~Eval.}} & LibriSpeech & Same-speaker & 548 & 449	& \numprint{997} \\ \cline{3-6}
  & test-clean & Different-speaker & \numprint{11196} & \numprint{9457} &	\numprint{20653} \\ \cline{2-6}
 & \multirow{2}{*}{VCTK-test} & Same-speaker & \numprint{2290} & \numprint{2096} & \numprint{4386} \\ \cline{3-6}
 % & & Same-speaker (common) & \numprint{346} & \numprint{354} & \numprint{700} \\  \cline{3-6}
 & & Different-speaker & \numprint{17894} & \numprint{18210} &\numprint{36104} \\ % \cline{3-6}
 % & & Different-speaker (common) & \numprint{4838} & \numprint{4952} & \numprint{9790}  \\
 \hline
  \end{tabular}}
\end{table}
\normalsize

Anonymisation performance is assessed objectively using an ASV system based on x-vector speaker embeddings and probabilistic linear discriminant analysis (PLDA) \cite{snyder2018x}.  
As shown in Figure~\ref{fig:asv-eval}, the EER is computed for a pair of evaluation scenarios~\cite{Tomashenko2021CSl,srivastava2021}:

\begin{itemize}\setlength\itemsep{0.05em}

\item \textit{Unprotected} --- No anonymisation (top of Figure~\ref{fig:asv-eval}).  Users share unprotected utterances (trial data).  The attacker attempts to identify the speaker using shared, unprotected trial data, unprotected enrolment data and an ASV system (denoted $ASV_\text{eval}$) trained using the original, unprotected \textit{LibriSpeech-train-clean-360} dataset.

\item \textit{Semi-informed} \cite{maouche2021enhancing} --- 
Users and the attacker apply anonymisation (bottom of Figure~\ref{fig:asv-eval}) to trial and enrolment data respectively.  
Users share anonymised trial data, to which the attacker also has access, in addition to unprotected enrollment data.  The attacker applies \textit{speaker-level} anonymisation to the latter using the same anonymisation system. 
Trial and enrollment utterances are, however, anonymised using different pseudo-speakers, since the attacker does not know the pseudo-speaker chosen by each user.
The attacker applies \textit{utterance-level} anonymisation\footnote{Previous work~\cite{shamsabadi2022dp} has shown that anonymising the training data at the utterance-level rather than the speaker-level results in a stronger attack.} to the \textit{LibriSpeech-train-clean-360} dataset and then trains a new ASV system, denoted $ASV_\text{eval}^{\text{anon}}$. 
Using this system, the attacker attempts to re-identify the original speaker of each trial utterance.
\end{itemize}

The EER for the unprotected scenario forms the baseline. 
The higher the EER for the semi-informed attack scenario, the better the privacy preservation.
The number of same-speaker and different-speaker trials in the development and evaluation datasets is given in Table~\ref{tab:trials}. For a given speaker and for both evaluation scenarios, all available enrolment utterances 
are used to compute an average enrolment x-vector.

\subsubsection{Utility metric: word error rate 
 (WER)}\label{sec:wer}
\begin{figure}[t!]
\centering\includegraphics[width=60mm]{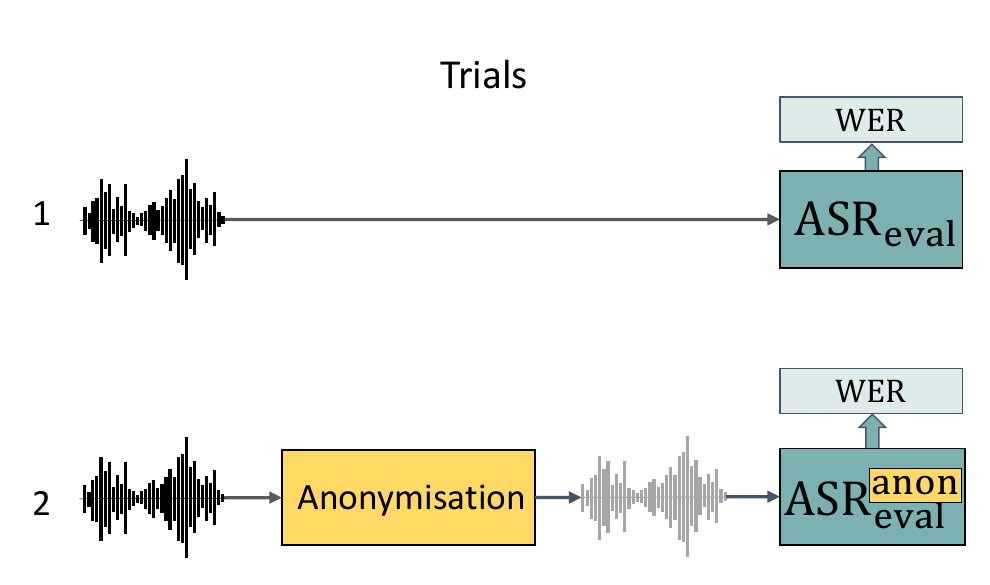}
\caption{ASR evaluation (1) original data decoded with \textrm{$ASR_\text{eval}$} trained on original data; 
(2)~\textit{speaker-level} anonymised data decoded with \textrm{$ASR_\text{eval}^\text{anon}$} trained on  \textit{utterance-level} anonymised  data.
The WER is computed on the \textit{trial} utterances of the development and evaluation datasets.}
\label{fig:asr-eval}
\end{figure}

The preservation of linguistic information is assessed objectively using an ASR system based on the Kaldi toolkit~\cite{povey2011kaldi}. 
We adapted the Kaldi recipe for \textit{LibriSpeech} to use an acoustic model with a factorised time delay neural network (TDNN-F) architecture~\cite{povey2018semi} and a \textit{large} trigram language model.
As shown in Figure~\ref{fig:asr-eval}, and as for the approach to gauge privacy, we consider two ASR evaluation scenarios: 
\begin{itemize}\setlength\itemsep{0.05em}
\item \textit{Unprotected} --- No anonymisation (top of Figure~\ref{fig:asr-eval}). Unprotected trial data is decoded using an ASR model (denoted \textrm{$ASR_\text{eval}$}) trained using the original \textit{LibriSpeech-train-clean-360} dataset.
\item \textit{Anonymised} --- Anonymised trial data is decoded using an ASR model (denoted \textrm{$ASR_\text{eval}^\text{anon}$}) trained using the \textit{LibriSpeech-train-clean-360} dataset after %the application of 
it is treated with
\textit{utterance-level} anonymisation, using the same anonymisation system as the trial data (bottom of Figure~\ref{fig:asr-eval}).
% application of (\textit{utterance-level}) anonymisation.
\end{itemize}

The first scenario again serves as a baseline.  
The lower the WER for the second, the better the utility preservation.
%obtained by the ASR model determines the level of utility (lower is better).

% \begin{figure}[t!]
% \centering\includegraphics[width=60mm]{Figures/ASR_ep_no_line.pdf}
% \caption{ASR evaluation (1) original data decoded with \textrm{$ASR_\text{eval}$} trained on original data; 
% (2)~\textit{speaker-level} anonymised data decoded with \textrm{$ASR_\text{eval}^\text{anon}$} trained on  \textit{utterance-level} anonymised  data.
% WER is computed on the \textit{trial} utterances of the development and evaluation datasets.}
% \label{fig:asr-eval}
% \end{figure}

\subsubsection{Privacy-utility tradeoff}
\label{sec:perf_objective}
New to the 2022 edition of VoicePrivacy was the use of multiple evaluation conditions.
These were introduced in recognition of the practical demand for different privacy-utility trade-offs and solutions which can be configured to operate at different operating points, as well as to provide \emph{common} optimisation criteria; for the 2020 challenge, participants had to select an appropriate privacy-utility trade-off themselves, resulting in each team essentially choosing \emph{different} optimisation criteria.
Evaluation conditions take the form of increasingly demanding minimum privacy requirements.
For each condition, systems which meet the corresponding minimum privacy condition are then ranked according to utility preservation.  
The primary privacy and utility metrics (EER and WER) are used for this purpose.

To stimulate progress, the 4 evaluation conditions are specified by a range of modest-to-ambitious minimum target EERs: 15\%, 20\%, 25\% and 30\%. 
%\{EER$_1$, \ldots, EER$_N$\}. 
Participants were encouraged to submit solutions to as many conditions as possible, with submissions to any one condition being required to achieve a weighted average EER for the evaluation set greater than the minimum target.
% EERs are weighted averages computed from those for the \textit{LibriSpeech-test-clean}, \textit{VCTK-test (common)} and \textit{VCTK-test (different)} datasets.
% Weights of 0.5, 0.1 and 0.4 respectively assign equal importance to LibriSpeech and VCTK datasets and account for the different number of trials in the two VCTK subsets.
EERs and WERs are equally-weighted averages computed from those for the \textit{LibriSpeech-test-clean} and \textit{VCTK-test} datasets
%, with weights being set in order to assign equal importance to each dataset 
(refer to~\cite{tomashenko2022voiceprivacy} for full details).
%The set of valid submissions for each minimum target EER were ranked according to the corresponding a
%WER results are equally-weighted averages computed from those for the \textit{LibriSpeech-test-clean} and \textit{VCTK-test} datasets. 
%The VoicePrivacy 2022 Challenge involves $N=4$ conditions with minimum target EERs of: EER$_1=15\%$, EER$_2=20\%$, EER$_3=25\%$, EER$_4=30\%$.
%
% Note, that for the EER threshold condition we consider an average EER over all three test sets.
%
%An example of system ranking according to this methodology is shown in Figure~\ref{fig:thresholds}.
%The lower the WER %\footnote{Note that, for the considered utility metric,  the lower the WER, the better the ASR word accuracy (calculated as $100$ -- WER), and the better the utility.}
%or a given EER condition, the better the rank of the considered system.
% A depiction of example results and system rankings according to this methodology is shown in Figure~\ref{fig:thresholds}.

\subsection{Secondary objective metrics}
Also new to VoicePrivacy 2022 was the introduction of a pair of secondary utility metrics, namely estimates of the pitch correlation $\pitchcorr$ and
the gain of voice distinctiveness $\gvd$.

\subsubsection{Pitch correlation  $\pitchcorr$}\label{subsubsec:f0}
Estimates of pitch correlation are used to approximate the degree to which an anonymisation system preserves intonation.
Following~\cite{Hirst07apraat}, the pitch correlation metric $\pitchcorr$ is 
the Pearson correlation between the pitch contours of original and anonymised utterances.  
The shortest of the two sequences is linearly interpolated so that its length matches that of the longest sequence.  
The temporal lag between original and anonymised utterances is then adjusted in order to maximise the Pearson cross-correlation when estimated using only segments during which both original and anonymised utterances are voiced. 
% We assume that the speech rate remains unchanged as a result of anonymisation.
%, so that the 
%Original and anonymised utterances are hence assumed to have similar length.\footnote{We setup a threshold of 65 ms for the maximum absolute length difference between an original and anonymised utterance.}
Estimates of $\pitchcorr$ %calculated for development and evaluation datasets, 
are averaged across the full set of utterances in a given data set.
%development and evaluation sets. separately.
%s of the pitch correlation values for all utterances in each dataset.
%
%%%%

%
%%%%
While a secondary metric, all submissions were required to achieve an average pitch correlation of $\pitchcorr>0.3$ for each dataset and for each evaluation condition to which a submission was made.
This threshold was set according to anonymisation results for baseline systems (described in Section~\ref{sec:baseline}).
%and a text-to-speech synthesis approach to anonymisation which does not preserve intonation.
%Solutions that achieve lower average pitch correlations will be considered invalid.

    \subsubsection{Gain of voice distinctiveness $\gvd$}\label{subsec:gainvd}
The gain of voice distinctiveness was adopted for VoicePrivacy 2022 to help observe the consistency in pseudo-voices for speaker-level anonymisation (see Section~\ref{sub:anon_task}).  
%aims to evaluate the requirement to preserve voice distinctiveness.
$\gvd$ is estimated using voice similarity matrices~\cite{noe2020speech,noe2021csl}.
A voice similarity matrix $M=( M(i,j))_{1 \le i \le N,1 \le j \le N}$ is defined for a set of $N$ speakers.  
$M(i,j)$ reflects the similarity between the voices of speakers $i$ and $j$:
%
% \begin{equation}
% \small
%      M(i,j) = \mathrm{sigmoid}\left({\frac{1}{n_{i}n_{j} }
%     \displaystyle\sum_{\substack{1 \le k \le n_{i} \text{ and } 1 \le l \le n_{j} \\ k\neq l \text{ if } n_j=n_j } }{\text{LLR}(x^{(i)}_{k},x^{(j)}_{l})}}\right)
%     \label{equ:: M}
% \end{equation}
% %
\begin{equation}
\small
     M(i,j) = \mathrm{sigmoid}\left({\frac{1}{n_{i}n_{j} }
    \displaystyle\sum_{\substack{1 \le k \le n_{i} \text{ and } 1 \le l \le n_{j} \\ k\neq l \text{ if } i=j } }{\text{LLR}(x^{(i)}_{k},x^{(j)}_{l})}}\right)
    \label{equ:: M}
\end{equation}
where $\text{LLR}(x^{(i)}_{k},x^{(j)}_{l})$ is the log-likelihood-ratio obtained by comparing the $k$-th utterance from the $i$-th speaker with the $l$-th utterance from the $j$-th speaker, and where $n_{i}$ and $n_j$ are the numbers of utterance for each speaker. 
LLRs are estimated using the $ASV_\text{eval}$ model trained using unprotected data.
Two matrices are computed:
$M_\text{oo}$, computed using unprotected utterances;
$M_\text{aa}$, computed using anonymised utterances.
The diagonal dominance $D_\text{diag}(M)$ is then computed for both. 
$D_\text{diag}(M)$ is the absolute difference between the mean values of diagonal and  off-diagonal elements:
\begin{equation}
\small
    D_{\text{diag}}(M)\hspace{-0.7mm}=\hspace{-0.7mm}\displaystyle
    \Bigg|
    \sum_{1\leq i \leq N} \frac{ M(i,i)}{N}
    \displaystyle
    - 
    \sum_{\substack{1 \le j \le N \text{ and } 1 \le k \le N \\j \neq k}}
    \frac{ M(j,k)}{N(N-1)}
    \Bigg|.
    \label{eq:ddiag}
\end{equation}
$\gvd$ is then computed as the ratio of diagonal dominance for each of the two matrices~\cite{noe2020speech}:
\begin{equation}
 \gvd = 10\log_{10}  \frac{D_\text{diag}(M_\text{aa})}{D_\text{diag}(M_\text{oo})}.   
\end{equation}

A gain of $\gvd=0$ implies the preservation of voice distinctiveness.  Positive and negative gains correspond respectively to an average increase or decrease in voice distinctiveness.

\subsection{Subjective assessment}

\begin{figure}[!t]
\centering
\includegraphics[width=0.95\linewidth]{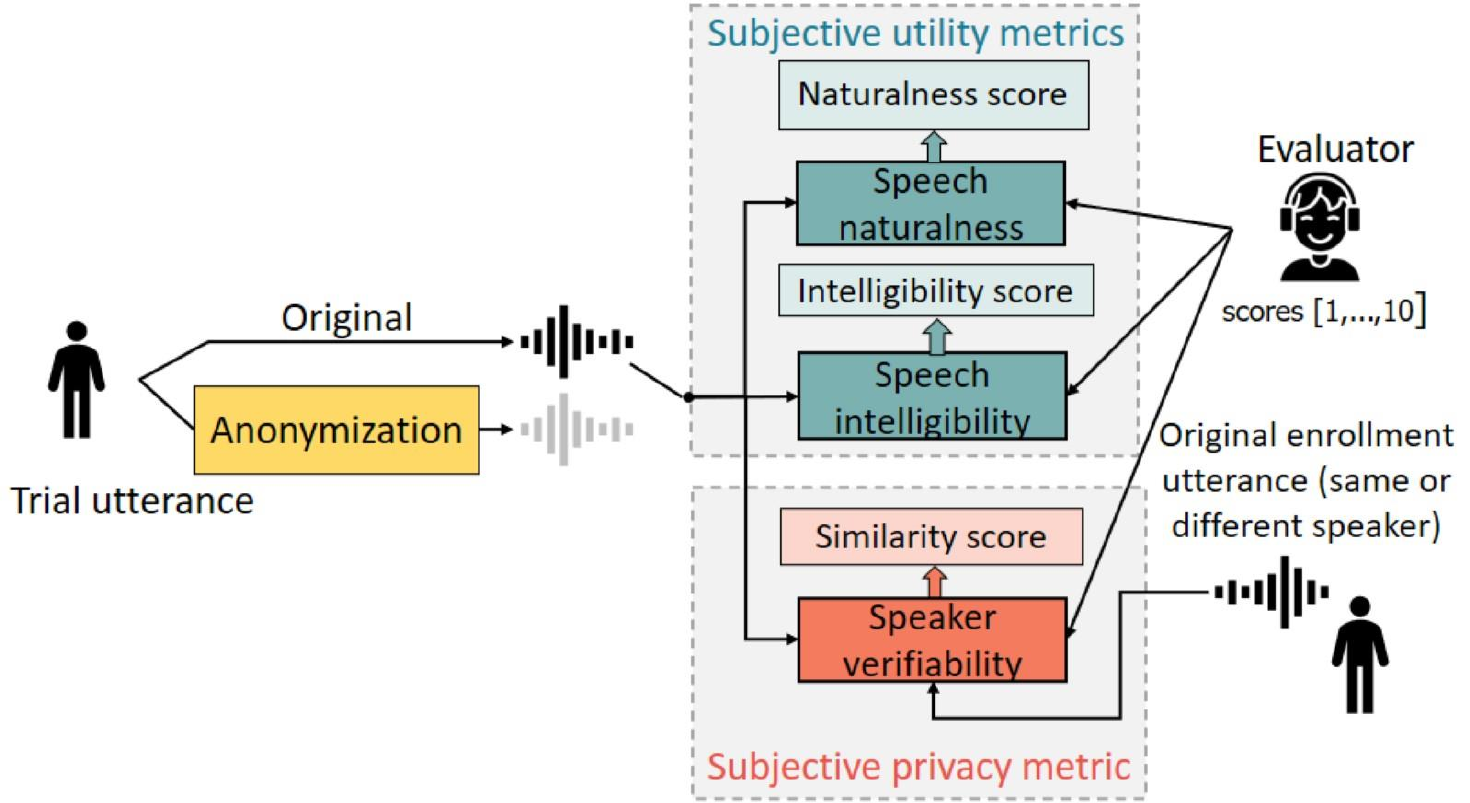}
\caption{The approach to subjective assessment for speech naturalness, intelligibility, and speaker verifiability.}
\label{fig:sub-eval-design}
\end{figure}

As illustrated in Figure~\ref{fig:sub-eval-design}, subjective metrics include estimates of speaker verifiability, speech intelligibility and speech naturalness.
Subjective evaluation tests were conducted by the challenge organisers.
For naturalness and intelligibility assessments, 
evaluators were asked to rate a \emph{single}
original or anonymised trial utterance at a time.
For naturalness, 
evaluators assigned a score from 1 (`totally unnatural') to 10 (`totally natural'). 
For intelligibility assessments, evaluators
assigned
a score from 1 (`totally unintelligible') to 10 (`totally intelligible').
\newparttwo{Both naturalness and intelligibility scores were normalised to within a range between 0 and 1 using rank normalisation~\cite{rosenberg2017bias},
with 0 representing the value of lowest naturalness/intelligibility, and 1 representing the highest.}
% respectively representing the scores of lowest and highest naturalness/intelligibility.}
Assessments of speaker verifiability were performed with \emph{pairs} of utterances, namely an unprotected enrolment utterance, and either an unprotected or anonymised trial utterance collected from the same or a different speaker. 
Evaluators assigned a speaker similarity score between 1 (`the trial and enrolment speakers are surely different') and 10 (`the trial and enrolment speaker are surely the same').
\newparttwo{Similarity scores were normalised in the same way as the naturalness and intelligibility scores.}

The evaluation trials are taken from the \textit{LibriSpeech-test-clean} dataset and include 1,352 unprotected utterances and 104 anonymised utterances per anonymisation system. 
Each subset of anonymised utterances contains 1 target trial and 1 non-target trial for each of 52 different speakers, evenly split between female and male.
The evaluation is performed by 52 native English speakers aged 18 to 70, of whom 40 where male, 11 were female, and 1 of undisclosed gender. Each evaluator rated 52 trials, 26 of which were unprotected, with the remaining utterance being anonymised either by a baseline or by one of the submitted systems.
With this configuration, all trial-enrolment pairs used for subjective evaluation were rated by at least one evaluator. % not sure if we want to point this out. @nick rephrase as you see fit

\section{Anonymisation systems}
\label{sec:system_description}

We describe the three VoicePrivacy 2022 baseline systems as well as those prepared by challenge participants.
A summary of the description is presented in Table~\ref{tab:submissions}.

\subsection{Baseline systems}\label{sec:baseline}

\begin{figure}[t]
\centering\includegraphics[width=87mm]{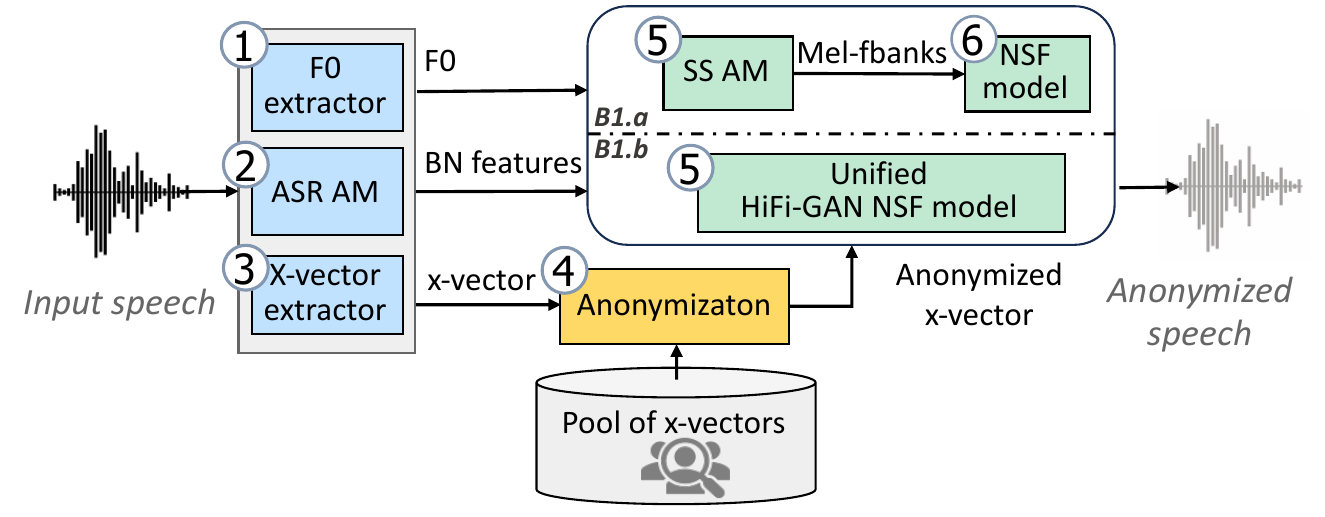}
\caption{Baseline anonymisation systems \texttt{B1.a} and \texttt{B1.b}.
}
\label{fig:baseline1a}
\end{figure}

Three different anonymisation systems were provided as challenge baselines, denoted \texttt{B1.a}, \texttt{B1.b}, and \texttt{B2}.
Baselines \texttt{B1.a} and \texttt{B1.b} are shown in Figure~\ref{fig:baseline1a}. 
Inspired by \cite{fang2019speaker}, \texttt{B1.a} uses x-vectors and neural waveform models, and comprises three steps: first,~x-vector~\cite{snyder2018x}, pitch~(F0) and bottleneck~(BN) features~\cite{povey2018semi} which encode linguistic content are extracted from the input utterance (blocks~\Circled{1}, \Circled{2}, \Circled{3}); second, the x-vector is anonymised (block~\Circled{4}); third, speech is synthesised using the anonymised x-vector and the original F0 and BN features (blocks~\Circled{5} and~\Circled{6}).

Pitch is estimated using YAAPT~\cite{YAAAAAPT}. BN features are 256-dimensional vectors extracted using a TDNN-F ASR acoustic model (AM)~\cite{povey2018semi}. Speaker encodings are 512-dimensional x-vectors extracted using a time-delay neural network (TDNN)~\cite{snyder2018x}.
%Block~\Circled{4} computes an 
% anonymised x-vectors are generated for every source x-vector by averaging a set of $N^*$ x-vectors selected at random from a larger set of $N$ x-vectors, itself composed of the $N$ farthest x-vectors, according to PLDA distances.\footnote{In the baseline, we use $N=200$ and $N^*=100$.}
The anonymisation function (yellow block in Figure~\ref{fig:baseline1a}) converts the original x-vector to an anonymised substitute.
Anonymised x-vectors are generated by averaging a set of $N^*$ x-vectors.  The latter are selected from a larger set of the $N$ farthest x-vectors from the original x-vector selected at random using a probabilistic linear discriminant analysis (PLDA)~\cite{plda} distance metric.
A speech synthesis (SS) AM generates Mel-filterbank features from the anonymised x-vector and F0+BN features.
The speech synthesis module is a neural source-filter (NSF) waveform model~\cite{wang2019neural}.
%synthesizes a speech signal from the anonymised x-vector, F0, and Mel-filterbank features.
Full details are available in~\cite{srivastava2020baseline}.
Baseline \texttt{B1.b} is the same as \texttt{B1.a}, except that the SS AM is removed and the NSF waveform model is fed directly with BN features. Moreover, the NSF is trained with an additional discriminator loss inspired by the HiFi-GAN system~\cite{kong2020hifi}.

Baseline \texttt{B2} is the technique presented in~\cite{patino2020speaker}, and is based purely on signal processing techniques.
The method utilises a coefficient $\alpha$ which is referred to as the \emph{McAdams} coefficient. 
Each pseudo-speaker is associated to a value of $\alpha$ randomly sampled from a uniform distribution within the range $(\alpha_{\text{min}}, \alpha_{\text{max}})$.
Linear predictive coding (LPC) is used to decompose the input utterance into a set of pole positions and an excitation signal. 
The poles positions are rotated within the complex plane to adjust the phase~$\phi$ to $\phi^{\alpha}$, hence shifting the formant positions of the input signal.
An anonymised utterance is then synthesised using the modified pole positions and the original excitation signal.

\begin{table*}[t]
%\resizebox{\linewidth}{!}{
% \begin{tblr}{l|p{3.5cm}p{3.5cm}p{3.5cm}p{3.5cm}}

\caption{A summary of the different systems, techniques and  performance.}
\label{tab:submissions}
% \begin{tblr}{X[1] |[1pt] X[3] |[gray] X[3] |[gray] X[3] |[1pt] X[3]}
\begin{tblr}{|[1pt] X[1] |[1pt] X[3] |[gray] X[3] |[gray] X[3] |[1pt] X[3] |[1pt]}
\hline[1pt]
\textbf{Team} & {\bf Feature extraction} & {\bf Anonymisation} & {\bf Resynthesis} & \textbf{Results summary} \\ \hline[1pt]
T04
    & {\scriptsize x-vectors and ECAPA vectors are concatenated to create speaker embeddings. Linguistic content is transcribed to phonemes. Removed F0 extraction.}
    & {\scriptsize Pseudo-speaker embeddings generated with GAN.}
    & {\scriptsize TTS model generates Mel-spectrograms that are converted to waveform by HiFi-GAN.}
    & {\scriptsize TTS-based approach provides excellent levels of privacy and utility, but barely passes the $\pitchcorr>0.3$ requirement.} \\
\hline
T11 
    & {\scriptsize Yingram for F0 extraction, U2++ for linguistic features. Speaker embeddings are one-hot representations of the speaker IDs encountered during training. One ``pseudo-speaker ID'' not corresponding to any real speaker is also stored.}
    & {\scriptsize Final pseudo-speaker embedding created by means of a weighted average between $K$ random speaker embeddings and the one-hot representation of the ``pseudo-speaker ID''.}
    & {\scriptsize HiFi-GAN is used to synthesise waveforms from AM-generated Mel-spectrograms.}
    & {\scriptsize \texttt{T11-p4} has one of the best results in terms of privacy and utility, but very low $\gvd$: all pseudo-speakers are ``similar''.}\\
\hline
T18
    & {\scriptsize As for \texttt{B1.a} and \texttt{B1.b}, except for their second system where x-vectors are replaced with Transformer-based ASR embeddings.}
    & {\scriptsize Two anonymisation strategies are proposed. The first uses adversarial noise to anonymise the speaker embedding. The second replaces the x-vector embedding with an ASR-based embedding.}
    & {\scriptsize As in \texttt{B1.a}.}
    & {\scriptsize Both approaches offer modest privacy improvement over \texttt{B1.a} and \texttt{B1.b} at the cost of reduced WER.} \\
\hline
T40
    & {\scriptsize F0 curve not extracted from the intput signal directly, but estimated from x-vector and BN features.}
    & {\scriptsize As in \texttt{B1.a} and \texttt{B1.b.}}
    & {\scriptsize As in \texttt{B1.b.}}
    & {\scriptsize Modest improvement over \texttt{B1} in terms of privacy.} \\
\hline[1pt]
T32
    & \SetCell[c=3]{c} {\scriptsize Signal processing-based approach: pitch shift with TD-PSOLA and PV-TSM.}
    &
    &
    & {\scriptsize Performance mostly on par with \texttt{B2} except for a higher $\pitchcorr$ and subjective intelligibility.} \\
\hline[1pt]
\end{tblr}
%}
\end{table*}

\subsection{Submitted systems}\label{sec:sys_participants}

The 2022 edition of the VoicePrivacy Challenge attracted submissions from 5 participating teams, all academic organisations.
For each evaluation condition, participants were required to designate one submission as their primary system with any others being designated as contrastive systems. Henceforth, submitted systems are named according to the following scheme: <team number>-<'p' if primary, 'c' if contrastive>-<incremental identifier>.
With full descriptions of each system available in the literature cited below, we report only brief summaries and provide an overview of the trends.

\textbf{Team T04}~\cite{T04} proposed an ASR+TTS-like approach.\footnote{\newpart{Code available at: \url{https://github.com/DigitalPhonetics/speaker-anonymization/tree/phonetic\_representations}}} A connectionist-temporal-classification/attention hybrid ASR model~\cite{ctc_attention} is used to transcribe input speech into phonemes, which are then converted to articulatory feature vectors~\cite{articulatory_fts}. Pseudo-speaker embeddings are created by means of a generative adversarial network (GAN)~\cite{gans}. Articulatory feature vectors and pseudo-speaker embeddings are used to synthesise spectro-temporal representations of an anonymised speech signal using the FastSpeech 2 TTS engine~\cite{fastspeech2}. A HiFi-GAN~\cite{kong2020hifi} is then used to synthesise speech signal outputs from the resulting representations. No pitch-related information from the original signal is used in the synthesis step.
A key motivation behind this approach is use of textual transcriptions in place of F0 curves to improve the
suppression of speaker-related information.

\textbf{Team T11}~\cite{T11} replaced the F0 trajectory with Yingrams~\cite{yingram}.
The authors argue that Yingram F0 extraction is more reliable in the case of
%this choice is motivated by the claim that regular F0 extraction algorithm might be imprecise in the presence of 
`creaky' voices caused by irregular glottal pulse periodicity.
%, and that Yingrams are more robust to such scenarios.
% to encode richer pitch information.
BN features are extracted with the U2++ model and WeNet toolkit~\cite{wenet}.
Unique to this system is the facility to configure the similarity between different pseudo-speaker voices, thereby controlling the privacy protection and voice distinctiveness trade-off.
Voice identity is encoded as a one-hot vector of size $N+1$, where $N$ is the number of speakers in the combined \textit{LibriTTS-train-clean-100} and \textit{LibriTTS-train-other-500} datasets. The additional vector component is referred to as a `pseudo-ID'. The pseudo-speaker representation is a weighted average between the one-hot vectors of $K$ random speakers and that of the pseudo-ID.
The weights act to control pseudo-speaker similarity.
A higher weight assigned to the pseudo-ID causes greater convergence of the pseudo-speakers to a single voice. 
Using pitch, bottleneck and speaker-related features, a spectrogram is generated using an acoustic model based on Tacotron 2~\cite{tacotron2}, and converted to a waveform using a HiFi-GAN vocoder~\cite{kong2020hifi}.

\textbf{Team T18}~\cite{T18} focused on enhancements to the anonymisation function of baseline \texttt{B1.a}.  
Two variations were investigated. 
The first uses adversarial noise to compute the pseudo-speaker embedding from the original speaker embedding.  The second explored the substitution of speaker embeddings with features extracted from the encoder component of a speech Transformer~\cite{speech_transformer}, with the rationale that they can, to some extent, encode speaker-related information.

\textbf{Team T40}~\cite{T40} proposed a variation of baseline \texttt{B1.b}. 
The YAAPT F0 extractor was replaced with a F0 regressor which uses BN features and pseudo-speaker embeddings to synthesise an F0 contour which better matches the voice identity of the pseudo-speaker. 
The hypothesis is that this approach can suppress speaker information contained in the F0 contour and result in more natural anonymised speech.

\textbf{Team T32}~\cite{T32} was the only team to adopt a signal-processing based approach. 
Motivated by the comparatively lightweight computational demands and since it does not require training data, they adopted a pitch shifting-based solution.  
Pitch shifting is achieved by either downsampling or upsampling before one of two different approaches is used to readjust the duration to that of the original speech signal. 
The first is based upon a time-domain pitch synchronous overlap-add (TD-PSOLA) approach~\cite{td-psola}.  The second uses phase-vocoder-based time-scale modification (PV-TSM)~\cite{pv-tsm}. 

% \begin{figure*}[!t]
%     \centering
%     \includesvg[width=\textwidth]{Figures/metrics_test4.svg}
%     \caption{Objective metrics results for the evaluation set.}
%     \label{fig:results_test}
% \end{figure*}

\begin{figure*}[!t]
    \centering
    \includegraphics[width=\textwidth]{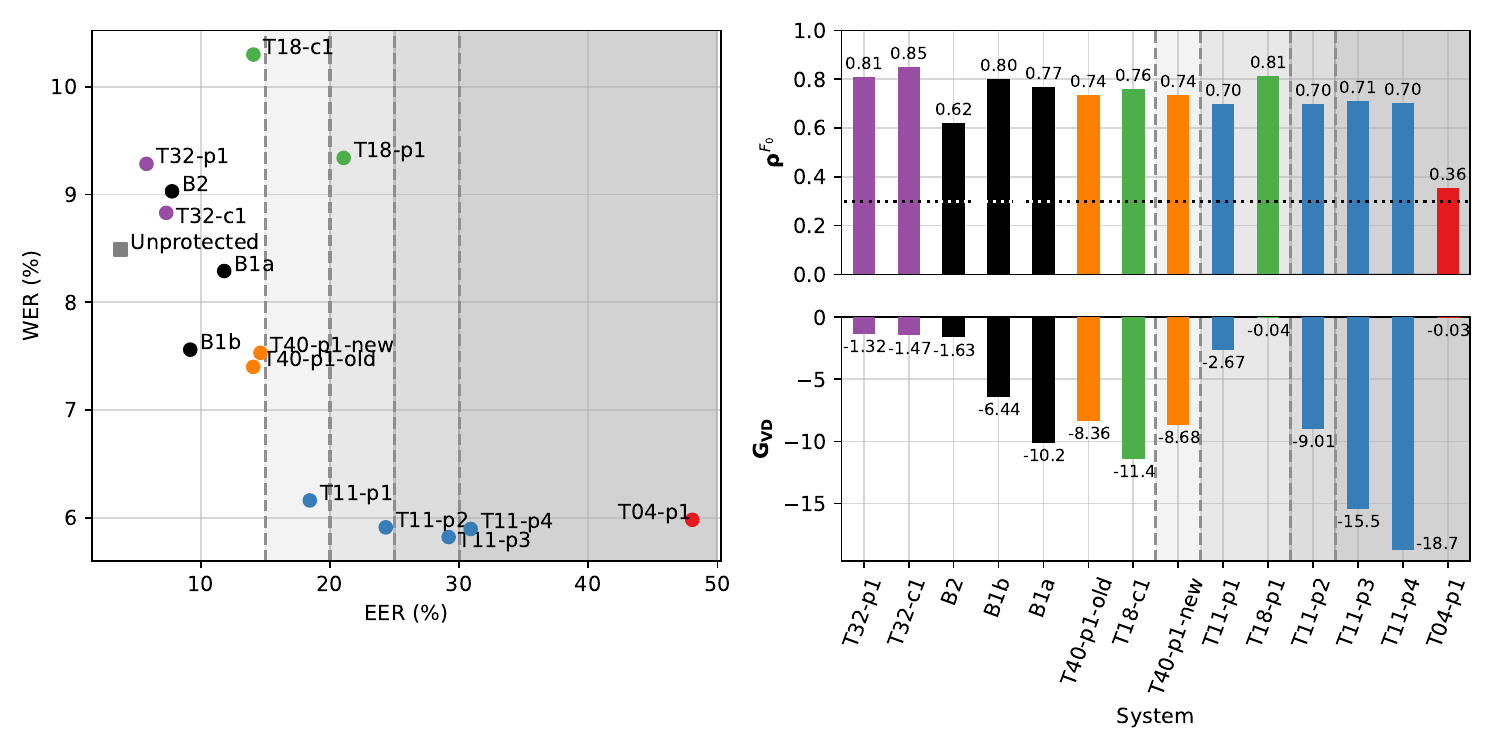}
    \caption{Objective evaluation results for the test set. Unprotected data was evaluated with $ASV_\text{eval}$ and $ASR_\text{eval}$, while anonymised data was evaluated with $ASV_\text{eval}^\text{anon}$ and $ASR_\text{eval}^\text{anon}$. Vertical dashed lines indicate the separation between different evaluation conditions. The horizontal dotted line in the pitch correlation plot shows the minimum pitch correlation threshold $\pitchcorr = 0.3$. Primary systems are denoted `-p' whereas the single contrastive system is denoted `-c'.}
    \label{fig:results_test}
\end{figure*}

% \subsection{Trends in speaker anonymisation}
\subsection{Trends}
%Most participants proposed variations of B1.a and B1.b. In Table~\ref{tab:submissions}, we provide 
A summary of the submitted systems is illustrated in Table~\ref{tab:submissions}.  In the following we describe the common trends and principal differences between them.
%techniques that are common how each submission differs from the baselines the three pipeline steps of feature extraction, anonymisation and resynthesis. Below is a brief discussion of each.

\textbf{Feature extraction -- } The majority of systems use the same three components used by the \texttt{B1.a} and \texttt{B1.b} baseline systems, namely 
%The idea of splitting the signal in the three elements of 
a speaker embedding, linguistic embeddings and a pitch contour.
%was mostly kept intact. 
While x-vectors remain popular, \texttt{T04} switched to the use of ECAPA-TDNN, as have a number of other works reported post-evaluation (see Section~\ref{sec:survey}).   
%still a widespread choice for encoding speaker identity, ECAPA vectors are growing in popularity, both among challenge participants and in speaker anonymisation research; generally, such a trend has taken place in other tasks involving speaker modeling [cite stuff].
Alternative approaches to the extraction of linguistic embeddings and F0 extraction include WeNet~\cite{wenet} and Yingram~\cite{yingram} respectively.

\textbf{Anonymisation -- } Probably because, instinctively, it has a major bearing on anonymisation performance, teams invested notable effort in improving the anonymisation function.
%the component which most As in the previous edition of the challenge, several approaches to identity anonymisation were proposed.  
% Common to the approach followed by \texttt{T04} and \texttt{T18} is an alternative to x-vector pooling, either with the use of a GAN or adversarial noise to generate anonymised speaker embeddings.
Common to the approaches of \texttt{T04}, \texttt{T11} and \texttt{T18} are alternatives to x-vector pooling using a GAN, a one-hot encoded speaker representation, or adversarial noise to generate anonymised speaker embeddings.
%An emerging trend is the attempt to circumvent the need to store an external pool of speakers, e.g. by generating new pseudo-speaker embeddings with a GAN.

\textbf{Resynthesis -- } There is comparatively little variation in the exploration of different approaches to resynthesis, perhaps indicating that most teams either found or expect the approach to synthesis to have comparatively less impact upon anonymisation performance.  Most x-vector--based systems used a HiFi-GAN (or a variation thereof), as for baseline \texttt{B1.b}.

\begin{figure*}
    \centering
    \begin{subfigure}[b]{0.95\textwidth}
        \includegraphics[width=0.9\linewidth]{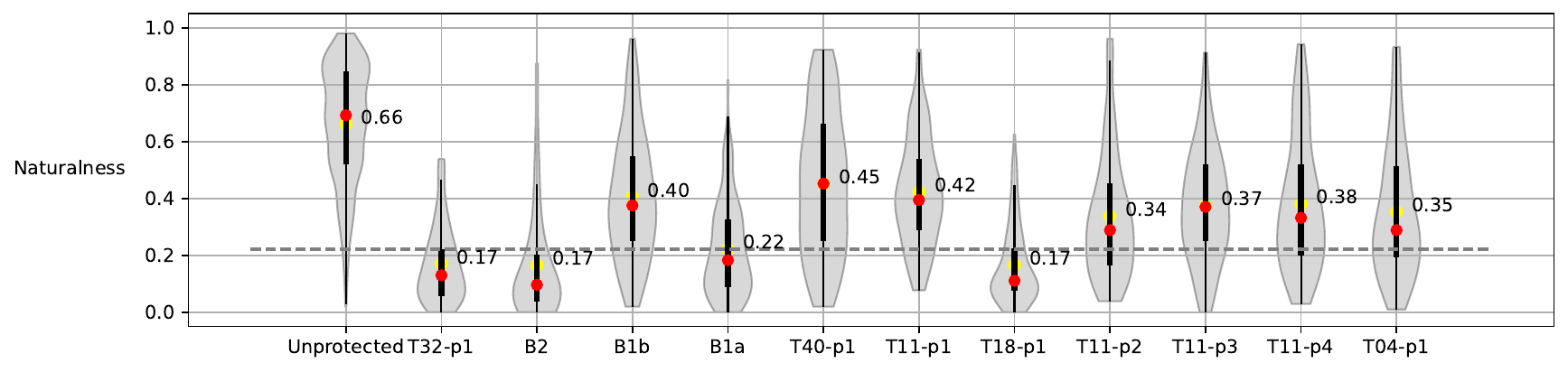}
        \caption{Distribution of naturalness scores}
        \label{subfig:naturalness}
    \end{subfigure}
    
    \vspace{0.4cm}
    
    \begin{subfigure}[b]{0.95\textwidth}
        \includegraphics[width=0.9\linewidth]{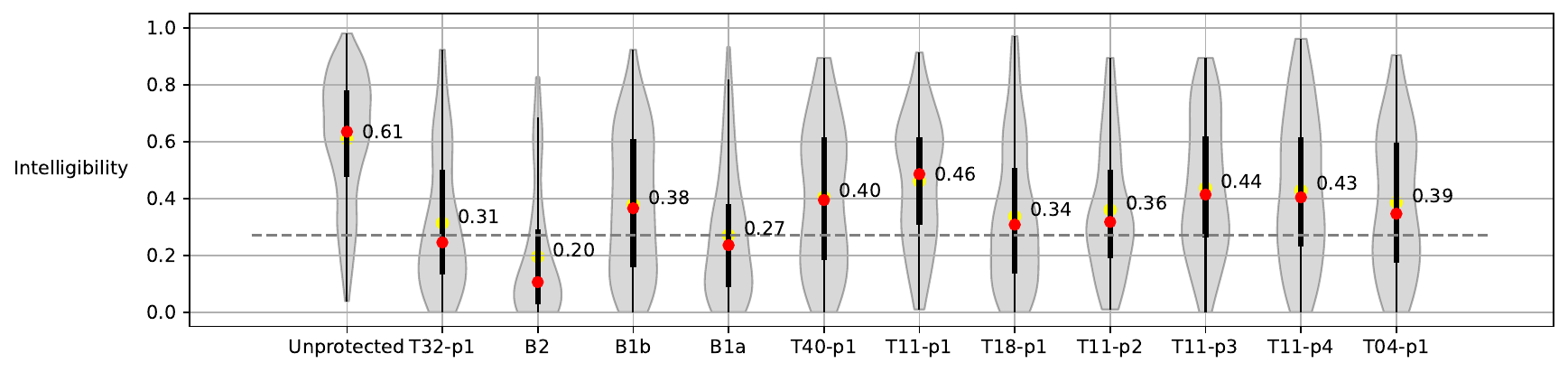}
        \caption{Distribution of intelligibility scores}
        \label{subfig:intelligibility}
    \end{subfigure}
    
    \caption{Subjective assessment results of baselines and submitted systems in terms of (a) naturalness and (b) intelligibility. Red and yellow dots represent respectively the median and the mean of each set of scores. The dashed gray line corresponds to the mean score of baseline \texttt{B1.a}, provided to facilitate the comparison with the other systems.
    }
    \label{fig:subjective_evaluation}
\end{figure*}

\section{Results}\label{sec:results}
We report results for baseline and submitted systems in terms of both primary and secondary objective metrics and subjective assessment.

% In this section we report the evaluation results for the systems described in Section~\ref{sec:system_description}. \added[id=rev, comment=1.8]{The results obtained as part of the challenge and those obtained as part of the post-evaluation analysis are both presented without distinction.}

\subsection{Objective evaluation results} \label{sec:obj_results}
Primary objective assessment results for baselines and all submitted systems for the evaluation set are illustrated in Figure~\ref{fig:results_test}. 
%For each system and each metric, we report the weighted average of the results obtained on the data partitions described in Section~\ref{sec:data}. The weights are chosen according to the size of the partitions, so that each contributes equally to the final result.
The plot to the left depicts the privacy-utility trade-off for unprotected data (grey points), for each baseline system (black points) and for each submission (coloured points).
All systems submitted by a given team are depicted by points of the same colour.
%For comparison, the privacy and utility values of the baselines (black dots) and the original data (gray dot) are reported. 
The vertical dashed bars depict the set of minimum target EERs of the four evaluation conditions defined in Section~\ref{sec:perf_objective}.

To the right of Figure~\ref{fig:results_test} are results for secondary metrics $\pitchcorr$ and $\gvd$. 
The set of systems is sorted according to the EER (low to high), and each bar has the same colour as corresponding points in the privacy-utility plots. 
The dashed bars again depict the juncture between evaluation conditions. 
The horizontal dotted line in the upper plot of pitch correlation results indicates the minimum threshold of $\pitchcorr=0.3$.

%Returning to the privacy-utility plots, 
All systems produce EERs higher than that for unprotected data and the majority of systems also outperform the baselines. 
The \texttt{T11-p3} submission produces the lowest WER for the 15\%, 20\% and 25\% minimum target EERs.  
The \texttt{T11-p4} system also achieves the lowest WER for the minimum target EER of 30\%.  
While these systems yield high pitch correlation values $\pitchcorr$, increased privacy (higher EERs) translates to a decrease in voice distinctiveness (lower $\gvd$), the lowest of all being for the \texttt{T11-p4} system.
These results nonetheless show the merit in the approach of Team \texttt{T11} to weight the pseudo-speaker embedding extraction in order to control the level of privacy.
Results show that greater privacy (higher EER) can be delivered with only negligible impact to utility (WER), albeit with notable degradation to voice distinctiveness ($\gvd$).

The EER for the \texttt{T04-p1} submission is substantially higher, just shy of a 50\% EER which would indicate perfect anonymisation.  
This can be attributed to the use of an ASR+TTS-like system.
The gain in voice distinctiveness, at $\gvd=-0.03$ is the best of all.
However, unsurprisingly for an ASR+TTS-like system, the pitch correlation is by far the lowest, with a value of $\pitchcorr=0.36$, albeit still above the threshold.
We return to this result later in \secref{sec:tts-pitch}.

The efforts of Team \texttt{T18} to improve the anonymisation function also appear to be successful.  Unfortunately, while the EER increases, the modifications also cause an increase in the WER, though the pitch correlation of $\pitchcorr=0.81$ and gain in voice distinctiveness $\gvd=-0.04$ are among the highest.

Perhaps due to the focus on modifications only to F0 extraction, the WER for Team \texttt{T40} submissions\footnote{Team \texttt{T40} submitted two versions their system: one before the challenge deadline, which contained an implementation error (named \texttt{T40-p1-old}), and one after the deadline, where the bug was fixed (\texttt{T40-p1-new}). While both systems are displayed in the objective metric results, only \texttt{T40-p1-old} was used in the subjective evaluation.} 
is relatively unaffected compared to results for the \texttt{B1.b} baseline from which their system is derived.
Increases to the EER are modest but the degradation to voice distinctiveness is relatively high.
While improvements were obtained for the development set, neither of the two approaches to pitch shifting explored by Team \texttt{T32} are successful in improving privacy in case of the evaluation set.  
The submissions of both Teams \texttt{T18} and \texttt{T32} show high pitch correlation and gain in voice distinctiveness.

\subsection{Subjective evaluation results}\label{sec:subj_results}
Results of subjective naturalness assessment are shown in Figure~\ref{subfig:naturalness}.
The first observation is a universal and substantial degradation in naturalness stemming from anonymisation. 
Baseline \texttt{B1.b} and derived \texttt{T40-p1} systems achieve among the highest scores and indicate that the NSF waveform model produces marginally more natural speech when fed directly with BN features instead of Mel-filterbank features.
%
%; it is surpassed only by 
%, which happens to be the most similar to B1b among all submitted systems,
%differing only in the pitch extraction part (the F0 curve is estimated from other input features, see Section~\ref{sec:sys_participants}).
%While we do not have sufficient data to precisely identify which element of the system contributes the most to making the produced speech sound natural, we do observe that the only deep learning-based system that achieves remarkably low naturalness (baseline B1a) is also the only one that does not employ some form of adversarial vocoder. The benefit of adversarial approaches to generative networks has been widely reported both inside and outside the audio domain~\cite{kong2020hifi,melgan}, and this result could be taken as a hint that such a benefit is also valid for the anonymisation task.
The relatively higher scores for Team \texttt{T11} systems and lower scores for \texttt{B1.a} and derived \texttt{T18-p1} systems suggest that adversarially trained vocoders produce more natural speech.
The \texttt{T04-p1} system, albeit ASR+TTS-like, is also competitive.
%, with scores being higher than that for the baseline B1.a system.
Naturalness scores for the \texttt{B2} baseline and \texttt{T32-p1} systems, both signal-processing based, are among the lowest.
%what is the dashed line?
%red is mean, yellow is median - can you check?
%why aren't the areas normalised to one?
%In any case, the gap in naturalness between original and anonymised speech is still quite wide, which calls for further research focus into this aspect.

%\subsubsection{Intelligibility}
The trends for intelligibility scores shown in Figure~\ref{subfig:intelligibility} reflect those for naturalness; 
%ow a similar trend to the naturalness scores: indeed, a Pearson correlation of $0.58$ is present between the two (see the rightmost plot of \figref{fig:wer_scatterplot}). % this was computed across all utterances, not on the means
%Despite its low $\boldsymbol{\rho}^{F_0}$ score, the TTS-based system \texttt{T04} does not seem to sound much less natural or intelligible to human listeners with respect to most other voice conversion-based approaches.
%Like for naturalness, 
anonymisation also universally degrades intelligibility. 
The \texttt{B1.b} baseline, the \texttt{T40-p1} and the set of \texttt{T11} systems are all competitive, as is the ASR+TTS-like \texttt{T04-p1} system. 
Scores for signal processing based solutions and the \texttt{B1.a} baseline are the lowest though, in contrast to results for objective utility assessment, the \texttt{T18-p1} system fares better.
We address the correlation between objective and subjective results later in \secref{sec:subj_vs_obj}.

\begin{figure*}
    \centering
    \includegraphics[width=\textwidth]{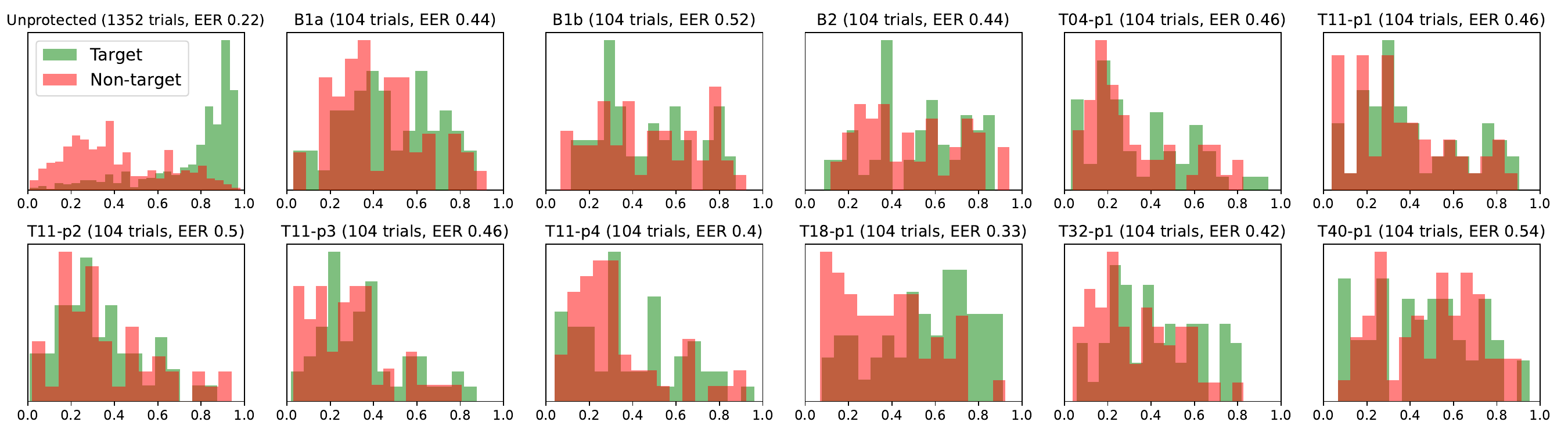}
    \caption{Distribution of target and non-target scores according to human listeners. For the top-left histogram, all data is unprotected.  For all others, the title above each histogram identities the system used to anonymise trial utterances. Enrolment utterances remain unprotected.}
    \label{fig:verif_distributions}
\end{figure*}

\begin{figure*}
    \centering
    \includegraphics[width=\textwidth]{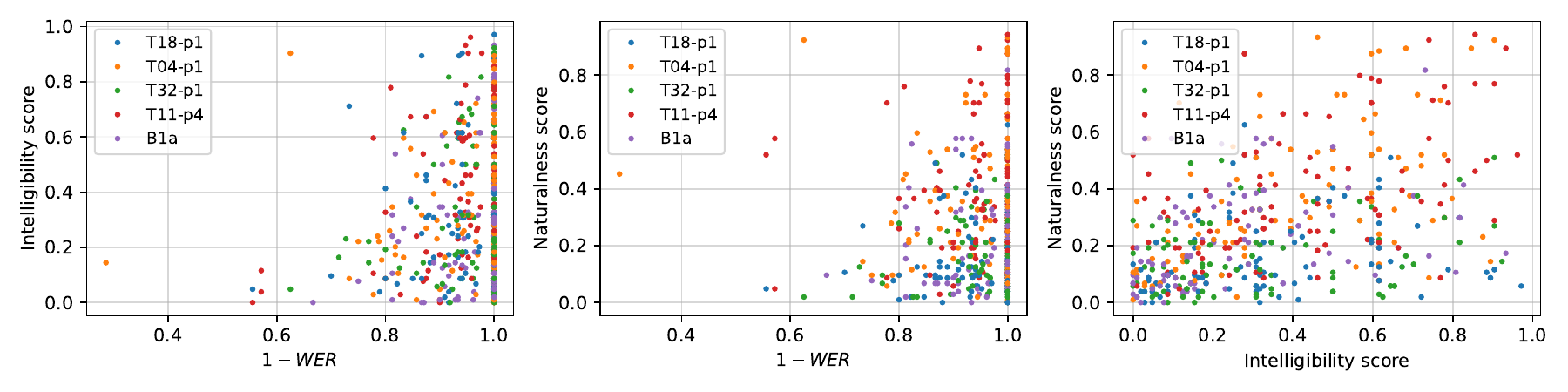}
    \caption{Scatterplots showing intelligibility score, naturalness score and the score corresponding to $1-\text{WER}$ for utterances of different systems.}
    \label{fig:wer_scatterplot}
\end{figure*}

% \begin{figure}
%     \centering
%     \includegraphics[width=0.85\columnwidth]{Figures/subjective/wer_subj_corr_small_v3.pdf}
%     \caption{Scatterplots showing intelligibility score, quality score and the score corresponding to $1-\text{WER}$ for utterances of different systems.}
%     \label{fig:wer_scatterplot}
% \end{figure}

%\subsubsection{Verifiability} \label{sec:subj_verif}
Verifiability score histograms shown in~\figref{fig:verif_distributions} 
%They are divided by anonymisation system, and the 
%Each histogram 
depict the distribution of verifiability scores for target trials (the speaker of both enrolment and trial utterances is the same) and non-target trials (the speaker of both enrolment and trial utterances is different).
The top-left-most histogram shows distributions for unprotected enrolment and trial utterances (no anonymisation)
% derived from 1352 trials % and 25 histogram bins
and that listeners can determine with reasonable reliability when the speaker of each utterance is the same or different.
%In such a case, some degree of separation is present between the two distributions, indicating that, on average, most of the target and non-target cases are correctly classified by human listeners.
All other histograms depict distributions 
% 104
when trial utterances are anonymised with one of 11 anonymisation baselines and submitted systems. 
Enrolment utterances remain unprotected.
The number of histogram bins is reduced compared to the plot for unprotected data because of the smaller number of trials. 
In almost all cases, the distributions for target and non-target trials are highly overlapping, indicating that listeners have greater difficulty to determine when the speaker of each utterance is the same or different.
%distributions almost completely overlap, indicating that the ability of the listener to distinguish between target and non-target trials is nearly reduced to a random guess. The only exception is \texttt{T18-p1}, where a slight degree of separation still emerges.
This includes distributions for the \texttt{B2} and \texttt{T32-p1} systems indicating that signal processing and deep learning based approaches to anonymisation are equally effective in the case of human listeners.
EERs estimated from the scores provided by human evaluators are almost all above 40\%. These estimates are, however, particularly noisy given the low number of trials, hence the reporting of histograms. % instead.

\section{Further analysis}\label{sec:further_analysis}
In this section, we report additional analyses performed post evaluation.  

\subsection{Subjective versus objective estimates of utility}\label{sec:subj_vs_obj}
The results in \figref{fig:results_test} show that WER estimates for some submissions are even lower than for unprotected data, implying that anonymisation actually \emph{improves objective estimates} of intelligibility (see Section~\ref{sec:improvedUtility} for a specific discussion of this issue).   
However, results in \figref{fig:subjective_evaluation}
show that anonymisation \emph{degrades subjective estimates} of intelligibility. 
It is hence of interest to explore these contradicting observations further.
%While they reflect performance according to different use cases, it is of interest to observe the correlation between subjective estimates of intelligibility and quality, and the objective WER. 
%We now focus on the relationship between the subjective intelligibility evaluation and the objective utility metric, the WER.
%For each utterance that was rated by human listeners, we consider its intelligibility score and utterance-wise WER. 

\figref{fig:wer_scatterplot} shows a set of scatter plots which depict the correlation between \emph{utterance-level} subjective intelligibility scores (left) and subjective naturalness scores (middle) against $1-\text{WER}$, for four different submissions and the baseline \texttt{B1.a} system.
%while the y coordinate is the subjective intelligibility score. 
%The second scatter plot is similar, with the y coordinate representing the subjective quality score.
In both cases, and for all five anonymisation systems, the correlation 
%between the objective and subjective metrics 
is low; the Pearson correlation with $1-\text{WER}$ is $0.14$ for intelligibility and $0.05$ for naturalness.
Curiously, for some utterances, $1-\text{WER} = 1$ (they are perfectly transcribed) but the corresponding subjective scores are near zero.
These observations suggest that objective and subjective measures are \emph{not functionally equivalent}, even though
%: rather, they reflect the viewpoints of human listeners and automated systems respectively, and are therefore complementary.
the WER is defined in~\cite{tomashenko2022voiceprivacy} as a proxy for intelligibility.
Based on this observation, it is clear that objective measures should no longer be considered as a proxy for subjective measures.
Even so, both are still of interest; they are indicators of anonymisation performance for different use cases, one involving the automatic treatment of anonymised utterances by machines and, for the other, consumption by human listeners.
\newparttwo{As shown in the scatterplot to the right of \figref{fig:wer_scatterplot}, there appears to be some degree of correlation between the two subjective measures, with their Pearson correlation coefficient being $0.58$. This might indicate that, from a perceptual perspective, the concept of `naturalness' is intrinsically linked to intelligibility. In the future, subjective metrics which better distinguish between the two aspects should be considered, e.g.\ assessments of intelligibility might be made by asking listeners to transcribe the spoken content of both anonymized and unprocessed utterances, and comparing the results.}

\subsection{ASR+TTS-based anonymisation}\label{sec:tts-pitch}

Since its inception, VoicePrivacy was designed to encourage the development of anonymisation systems which preserve linguistic and para-linguistic speech attributes. 
%such as intonation (in addition to linguistic attributes).
%It can then be assumed that voice conversion-based systems have better potential than TTS-based systems; the former are more likely than the latter to preserve para-linguistic attributes.
Despite it being a design goal, a means to gauge the preservation of para-linguistic attributes was missing in 2020.  
This was an obvious weakness since ASR systems could be used to generate an intermediate transcription of the input before the application of TTS to generate  perfectly voice-anonymised and intelligible utterances, albeit without the para-linguistic attributes of the input.
%voice conversion-based and related
While the submission of ASR+TTS systems was not prohibited, the pitch correlation metric $\pitchcorr$ and minimum threshold
were hence introduced for the VoicePrivacy 2022 edition in order to favour solutions (e.g.\ those based on voice conversion) which offer better potential for anonymisation while also preserving para-linguistic attributes. % at least to some extent. 
%TTS-based solutions were, however, not prohibited, thereby leaving opportunity to explore TTS-based solutions 
%was introduced in the  VoicePrivacy 2022 challenge in order to encourage the development of 
%The 2022 edition saw the introduction of and the 

% \textbf{Legitimize text-to-speech?}
The majority of teams pursued voice conversion-based solutions.  Team \texttt{T04} was alone in exploring an ASR+TTS-like solution which, perhaps unsurprisingly, delivers  near-to-perfect objective anonymisation results
%with a EER of close to $50\%$ 
and among the lowest WERs for test data (see \figref{fig:results_test}), as well as competitive subjective naturalness and intelligibility assessment results (see \figref{fig:subjective_evaluation}).
%It is likely that such a result was achieved thanks to the use of a TTS-based system that generates an intermediate transcription of the linguistic content and re-synthesizes a waveform from it.
The gain in voice distinctiveness $\gvd$ is the best of all and, interestingly, the pitch correlation $\pitchcorr$ also exceeds the minimum threshold.
%, barely fulfilling the requirement of $\boldsymbol{\rho}^{F_0}>0.3$, seems to achieve relatively good results on all other evaluation metrics, including 
%subjective ones (see Section~\ref{sec:subj_results}).
%We conclude that, if prosody preservation is not a strict requirement, 
Whether ASR+TTS solutions are appropriate, whether the minimum pitch correlation threshold for VoicePrivacy 2022 was perhaps too low, or even whether pitch correlation is sufficient on its own as a measure of para-linguistic attribute preservation, are all matters of opinion. 
Ultimately, all are also dependent upon the specific use case scenario.  
%Undeniably, though, 
Given the promising EER, WER and $\gvd$, ASR+TTS systems warrant continued attention, especially given that various techniques could be applied readily to boost the pitch correlation or the preservation of para-linguistic attributes in order to improve their competitiveness with voice conversion-based solutions.
In the following, we present some initial work designed to gauge the potential.

%-based systems may constitute a sound approach to speaker anonymisation, and should not be ruled out a-priori.

%Moreover, even with a strict pitch correlation requirement, we argue that it might still be possible to make TTS systems viable with additional alignment techniques. In the following, we support our claim by presenting a preliminary investigation about how TTS-generated F0 curves can be re-aligned to those extracted from the the original utterances.

% \begin{figure*}
%     \centering
%     \includegraphics[width=\linewidth]{Figures/84-121123-0000_DTW.pdf}
%     \caption{F0 curves and pitch correlation of utterance 84-121123-0000 (from LibriSpeech test set) before and after DTW. Top right: F0 curves of the original, T04-anonymised, and T11-anonymised utterance. Top left: F0 curves of original and T11-anonymised utterances after DTW. Bottom left: F0 curves of original and T04-anonymised utterances after DTW. Bottom right: pitch correlation values of the utterances before and after DTW.}
%     \label{fig:dtw_1}
% \end{figure*}

\begin{figure} 
    \centering
    \includegraphics[width=\columnwidth]{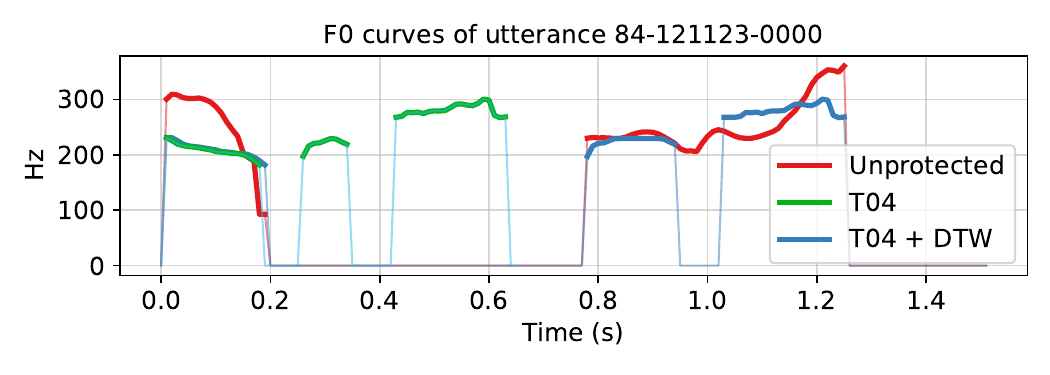}
    \caption{F0 contours for utterance 84-121123-0000 (from the \textit{LibriSpeech-test-clean} set). Red -- unprotected; green -- anonymised with \texttt{T04} system; blue -- anonymised with \texttt{T04} system and then aligned to the original utterance using dynamic time warping (DTW).}
    \label{fig:dtw_1}
\end{figure}

\subsection{Pitch realignment}

We applied a trivial form of pitch realignment using dynamic time warping (DTW) to determine whether the pitch correlation for ASR+TTS anonymised utterances can be improved, and hence whether even higher minimum thresholds might also be met with ASR+TTS approaches.  
Whereas the pitch correlation $\pitchcorr$ is estimated by compensating for any misalignment between anonymised and unprotected utterances using linear warping functions, DTW supports more flexible non-linear warping functions. 
This greater flexibility should result in improved pitch correlation.
We set the DTW local continuity constraints to warp the pitch correlation of anonymised utterances onto those of corresponding unprotected utterances.
%; the latter hence remain untouched.

Figure~\ref{fig:dtw_1} shows three pitch contours for the same expressive utterance \emph{``Go! Do you hear?''}, contained within the \textit{LibriSpeech} test set.  
They correspond to:
the unprotected utterance (red); the \texttt{T04}-anonymised utterance (green); the corresponding DTW-aligned utterance (blue).
Alignment is shown to improve greatly with the application of DTW.

We applied the same pitch realignment process to the full test set and utterances treated with one of the six systems shown in Table~\ref{tab:dtw_2}.
Pitch correlation results, shown in all cases with and without the application of DTW alignment, improve markedly for all systems, and for the \texttt{T04-p1} system the most.  
The improvement in this case is substantial, giving values of $\pitchcorr$ higher than those for all original systems without DTW alignment.  
Still, the result of $\pitchcorr =0.83$ is lower than that for all systems with DTW alignment.
This result suggests that high values of $\pitchcorr$ can be obtained with minimal additional effort, and either that the threshold of 0.3 is far too low, and/or the metric itself is deficient. 
We stress that we did not use the warped F0 contours to resynthesise speech signals, the naturalness and intelligibility of which would inevitably degrade.
Resynthesis would require further work to warp-adjust articulatory or linguistic features, or an alternative approach to DTW, e.g.\ by operating upon spectral features.  

\begin{table}[]
    \centering
    \begin{tblr}{|[1pt] c|c|c|[1pt]}
    \hline[1pt]
    &  $\pitchcorr$ \textbf{w/o DTW} & $\pitchcorr$ \textbf{with DTW}\\
    \hline[1pt]
    T04-p1 & 0.36 & 0.83 \\
    \hline
    T11-p4 & 0.70 & 0.91 \\
    \hline
    T18-p1 & 0.70 & 0.93 \\
    \hline
    T32-p1 & 0.81 & 0.94 \\
    \hline
    T40-p1 (old) & 0.74 & 0.90 \\
    \hline
    T40-p1 (new) & 0.74 & 0.90 \\
    \hline[1pt]
    \end{tblr}
    \caption{Pitch correlation values of different anonymisation systems with and without the application of DTW.}
    \label{tab:dtw_2}
\end{table}

\begin{comment}
\subsection{Privacy protection and voice distinctiveness}
% \textbf{Anti-correlation between distinctiveness and privacy.}
System T11 includes a parameter $\alpha$ that tunes the privacy protection level:
% Team T11's model includes a parameter $\alpha$ that allows to tune the privacy level of the system:
the higher $\alpha$ is, the more all pseudo-speaker embeddings tend to converge to one single point, increasing privacy at the expense of voice distinctiveness. This result suggests that, if distinguishing individual speakers is a strict requirement, mapping all utterances to a single pseudo-speaker might represent a straightforward, yet effective approach to enhance privacy preservation.

It should be noted that, in single--pseudo-speaker anonymisation systems, the pseudo-speaker will also be known by a potential privacy adversary, making them a completely \emph{informed} attacker.
This anonymisation setup was more thoroughly evaluated in~\cite{champion21_whitebox}, where it is referred to as a \emph{constant target strategy}.
In~\cite{champion_thesis}, the same authors show that the pseudo-speaker selection method can bias the attacker into training a sub-optimal $ASV_{eval}^{anon}$ model; they then suggest that
% using the same pseudo-speaker for every speaker
using the \emph{constant target strategy} during evaluation
would allow to measure the erasure of personal information contained in the signal independently of the choice of the pseudo-speaker, thus allowing for a more accurate estimation of privacy protection.

\end{comment}

\subsection{Utility increase with anonymisation}
\label{sec:improvedUtility}

The plot to the left in \figref{fig:results_test} shows that some anonymisation systems lead to lower WERs than that for unprotected data.
This would suggest that, contrary to intuition,  
%results shown in Several systems appear to improve upon the original unprotected test WER of $8.5\%$: B1b and T40 score around $7.5\%$, T04 and T11 around $6\%$. In the development set results (not shown here), only T04 maintain this property, with a slightly smaller gap ($7.3\%$ WER for unprotected data, $6.6\%$ for T04, around $5.6\%$ for T11).
%In other words, it appears that 
%use of the $ASR_{eval}^{anon}$ system trained on similarly anonymised data, 
anonymisation \emph{improves} intelligibility.
%similarly trained anonymisation by these systems appears to make the signals more intelligible to the $ASR_{eval}^{anon}$ model.
%However, when interpreting these results, 
%
%Two aspects should be considered when interpreting this observation.
%First, $ASR_{eval}^{anon}$ models are trained from scratch on speech anonymised by the same system that anonymised their test data.
%This increases the risk of overfitting on a specific system, especially those with low $G_\text{VD}$ that produce abnormally similar pseudo-speakers.
%Moreover, from a practical standpoint, this scenario implies that the entity that makes use of the speech data for the downstream task of ASR is aware that the data is anonymised and which system was used to anonymise it; this might be unrealistic and even undesirable, since a more knowledgeable privacy adversary is also generally more powerful~\cite{srivastava2021, panariello_vocoder_2023}.
%
This would be a rather favourable and unrealistic interpretation.
%Second, 

The $ASR_{eval}$ model is trained using unprotected data 
%the training data of 
%$ASR_{eval}$ only includes 
sourced from the \emph{LibriSpeech-train-clean-360} dataset. 
Conversely, the $ASR_\text{eval}^\text{anon}$ model is trained using the anonymised version of the same dataset. 
The anonymisation system, however, is trained 
%, which has been processed by a further model (the anonymisation system) that was trained on 
using a much larger amount of data, namely that sourced from the
%. Namely, according to the evaluation plan~\cite{tomashenko2022voiceprivacy}, the anonymisation system can be trained on 
\emph{LibriSpeech-train-clean-100}, \emph{LibriSpeech-train-other-500}, and \emph{VoxCeleb 1-2} datasets.  
This likely leads to the leaking of anonymisation training data into the $ASR_\text{eval}^\text{anon}$ model, implying that it is exposed to more data during training than the 
$ASR_{eval}$ model.
While comparisons made between two different anonymisation systems involve models trained under identical data conditions, comparisons in utility before and after anonymisation do not. 
Results should then be interpreted with appropriate caution. 
%This would explain why some systems seemingly improve the utility of the original data. 
Fairer comparisons of utility before and after anonymisation might be made if the data used for the training of the the $ASR_{eval}$ model were augmented with the same data used in the training of anonymisation systems.  
It should be noted, though, that this would lead to other undesirable issues concerning data overlap.

\section{Post-evaluation work and results}\label{sec:survey}

%While the VoicePrivacy Challenge aims to foster the development of novel and more effective speaker anonymisation algorithms, the field has of course progressed even outside of it. % does this sound lame?
%For the sake of completeness, 

In this section we provide an overview of relevant, peer-reviewed work published in the open literature post evaluation.
%in the the main findings in speaker anonymisation published since the end of the 2022 edition of the Challenge.

% x-vector based
%\textbf{X-vector--based pipelines} remain the most popular approach and have been pursued further in~\cite{miao_ohnn_2023,pierre_are_2022,panariello_vocoder_202}.  
Improvements to the \textbf{anonymisation function} based on \emph{orthogonal Householder neural networks} (OHNNs) are reported in~\cite{miao_ohnn_2023}.  
The layers of the OHNN consist in a linear transformation defined by an orthogonal matrix. 
The parameters of each layer are trained to maximise the distance between original and anonymised speaker embeddings while preserving voice distinctiveness and the overall distribution in the embedding space. 
Use of OHNNs results in greatly improved anonymisation performance to nearly $50\%$ EER in a semi-informed attack scenario without loss to utility and is hence an attractive direction for future work.
\newpart{To the best of our knowledge,~\cite{miao_ohnn_2023} is also the only post-evaluation work which investigates the effect of anonymising \mbox{non-English} speech using models trained on mostly English corpora.}

\textbf{Vocoder contributions to anonymisation}, discussed in~\cite{panariello_vocoder_2023}, can even outweigh those of the anonymisation function;
%shows that .   showed that tcontributions to anonymisation performance come from both the anonymisation the vocoder contributes to anonymisation itself actively contributes to the anonymisation process: 
the x-vector extracted from the anonymised utterance often differs substantially from the x-vector at the vocoder input.  This phenomenon, referred to as \emph{vocoder drift}, likely stems from the leakage and re-entanglement of speaker information contained in linguistic and prosodic features~\cite{panariello23_spsc}. 
It is argued that poor control of the speaker embedding space can hinder the development of better anonymisation functions, implying that future work should also take vocoder effects into account.
%contributes to enhancing anonymisation by making the final x-vector position more unpredictable, and also appears to be linked with speaker information leakage in linguistic and prosodic features.

\textbf{Privacy leakage} was reported earlier in~\cite{pierre_are_2022} which shows that personally identifiable information contained in sources other than the speaker embedding can leak into anonymised speech signals.  
The authors show that leakage can be reduced through the application of vector quantisation to the linguistic features and argue that
%leads to higher levels of privacy preservation, suggesting that some degree of speaker information is captured by the BN extractor. 
%The authors claim that 
properly \emph{disentangled} features are key to achieving more effective anonymisation. The EER of their baseline system sees an increase of EER from $8\%$ to $16\%$ when adding a quantization bottleneck on linguistic features, at the cost of a WER degradation from $7\%$ to $10\%$.
More elaborate disentanglement techniques, e.g.\ to isolate cues relating to the speaker sex~\cite{noe_2023_hiding}, have been reported.
These techniques will likely attract greater attention in the future; they might one day allow
%They have potential to extend anonymisation towards more flexible privacy preservation frameworks with which 
for multiple privacy-sensitive attributes (e.g.\ the voice identity and the speaker's sex and age) to be selectively and simultaneously suppressed or manipulated. 

The \textbf{pitch correlation of ASR+TTS-like anonymisation} was explored further by the \texttt{T04} participants.
%, which was proposed as an enhancement to participant system T04. In line with our observations in \secref{sec:obj_results}, 
They report~\cite{meyer_prosody_2023} a technique to transfer prosodic information contained in the input utterance to the synthesised, anonymised output. 
%in a TTS-based anonymisation system.
%They start from a TTS model that normally predicts three values for each input phoneme: duration, average pitch and average energy (i.e. L2 norm of the amplitudes of an STFT frame). 
They explored the extraction of phoneme duration, average pitch and average energy, 
%(i.e.\ L2 norm of the amplitudes of an STFT frame), 
not from phoneme transcriptions as in their original approach~\cite{T04} but, instead, directly from
%modify the system so that these values are estimated directly from the 
the input utterance.
While pitch correlation is shown to improve from 0.3 (for the original \texttt{T04} system) to ~0.7, the experiments were conducted with a lazy-informed attack scenario instead of the VoicePrivacy semi-informed attack scenario.

\textbf{Alternatives to x-vector speaker embeddings} have also been investigated.
%Other approaches to speaker anonymisation have also surfaced, employing approaches and architectures that differ from the usual x-vector--based pipelines by various degrees.
A flow-based TTS model, adapted to perform voice conversion, is reported in~\cite{nespoli_two-stage_2023}.
The spectrogram of the original utterance is passed through an encoder and a direct flow model conditioned on the original speaker embedding.
The resulting latent utterance representation is then fed into an inverse flow model conditioned on a pseudo-speaker embedding and an output waveform is resynthesised. 
A competitive EER of 22\% is achieved with minimal utility degradation.
%when using the author's re-implementation of the evaluation pipeline.
%using a vocoder.
%While this alone provides a decent privacy protection, the authors experimented with passing the anonymised utterance through a second anonymisation system (baseline B1a), enhancing the privacy protection level while trading off some utility. They also report that processing the utterance with B1a first and the flow-based model afterwards results in a greater utility preservation, but inferior privacy protection.
%The authors of~\cite{panariello_speaker_2023} also proposed a different 
An approach to speaker anonymisation based on auto-regressive modelling of neural audio codecs is reported in~\cite{panariello_speaker_2023}. 
An input utterance is converted to a set of semantic and acoustic tokens using a semantic encoder and a neural audio codec respectively, and a transfomer model is then trained to predict the acoustic tokens from the semantic tokens.
Voice conversion is then performed by conditioning the Transformer on acoustic tokens extracted from a different speaker.
% Both methods improve 
The method gives an EER of 32\% and both pitch correlation and voice distinctiveness are relatively well preserved.
%they both result in come at the achieves good privacy protection levels, 
Nonetheless, the methods in~\cite{nespoli_two-stage_2023} and~\cite{panariello_speaker_2023} both degrade utility.

The authors of system \texttt{T11} proposed a variation of baseline \texttt{B1.a} for which the x-vector speaker encoder is replaced by a formant-based identity representation~\cite{yao_distinguishable_2023}. 
%The work is a follow-up of the system submitted to the challenge, which suffered from low voice distinctiveness. 
BN features are computed in the same way as for the original \texttt{T11} system. 
The first five formants are extracted from the input utterance, and further processed by different neural-network based encoders to create a speaker representation that better preserves voice distinctiveness. 
F0 curves are extracted with PRAAT and the final waveform is synthesised with an acoustic model and a vocoder. 
The speaker identity is anonymised using a variety of techniques to scaling the F0 and formant position. 
The use of larger scaling values can trade stronger  anonymisation for degraded utility, pitch correlation and, still, voice distinctiveness.
% this one has multiple scaling values and i don't know which one to report, give me a second
The configuration which achieves the best voice distinctiveness ($\gvd \approx -4.5$) yields an EER of 21\% with utility comparable to that of the original \texttt{T11} system.

\section{Future directions}
%We might acknowledge that, perhaps with the exception of one recent work, we remain far from real anonymisation.
%Since we remain far from an ideal solution to voice anonymisation, 
A third edition of the VoicePrivacy challenge is planned.
In the following we discuss some of the priorities and likely directions.

The choice of \textbf{source data} has been queried within the community.
\textit{LibriSpeech} and \textit{VCTK} datasets contain relatively high-quality speech data \emph{read} from book chapters and newspaper text.
Read speech lacks the spontaneity of speech collected in settings which might be more representative of practical use case scenarios.
\textit{VCTK} data was collected with a single microphone and from within a single, hemi-anechoic chamber.
%While collection conditions for \textit{LibriSpeech} data were comparatively uncontrolled, 
In contrast, it can be assumed reasonably that, for a given speaker, \textit{LibriSpeech} data were likely collected using the same microphone and from within the same room.  
Its use for ASV experiments is hence potentially problematic since the use of cues related to the voice as well as the channel characteristics can be used for speaker verification.
The impact of persisting channel characteristics and their potential use to infer speaker identity remains untested in the scope of VoicePrivacy and merits attention in the design of future challenges. 
There is also concern that the size of the datasets used for ASV training (being vastly smaller than more popular and current alternatives) is insufficient, that the resulting models are hence poorly trained, and that the use of larger datasets will influence anonymisation performance, leading to different findings.

%However, speaker identity could be reconstructed from other factors, such as spoken content containing PII or recording conditions. 
%Moreover, such a simple setting might ease the re-synthesis process and result in overly optimistic WER estimates.
%The choice of evaluation datasets in the next challenge edition should attempt to reflect more varied and realistic scenarios.
% Use of LibriSpeech for ASV - commonvoice?  Training and testing using LibriSpeech makes both rather poor - we can obtain the same performance using fewer utterance which means the training / ASV system isn't really working well.  Are we even recognising the speaker, or are we recognising the book or the *channel*, the speaking style, etc.?
% For measuring WER / utility, LibriSpeech is also too simple - it lacks spontaneity that might characterise practical applications.

% 1) Pierre's view on: 
% The requirement for \textbf{speaker-level anonymisation} (see Section~\ref{sub:anon_task}) also attracted feedback.
%has seen much-diverging opinions from the community, ranging from participants needing help understanding why it was a primary requirement to them having trouble making mechanistic conclusions from the $G_{\text{VD}}$ metric.
%The requirement (d) from Section \ref{sub:anon_task} about 
%was imposed for this evaluation purpose.
%However, beyond the factors mentioned earlier, there is a vulnerability in this requirement: 
\textbf{Speaker-level anonymisation} (see Section~\ref{sub:anon_task}) facilitates its application to multi-speaker conversations.
Nonetheless, %some use cases may not require it.
% Some participants found this to be an unwelcome complication, advocating instead for 
utterance-level anonymisation %and single-speaker use case. 
may be sufficient for other applications.
Assessment is also arguably inconsistent with speaker-level anonymisation:
performance is assessed at the utterance-level and does not take into account the advantage that an attacker can gain from knowing that anonymisation is applied at the speaker-level.  
%It is a reasonable assumption that the level of protection for one utterance might be different to that of another.
Given that each utterance corresponding to the same speaker is anonymised with the same pseudo-voice, an attacker need only overcome the protection of one (perhaps weakly anonymised) utterance and then link utterances having the same pseudo-voice in order to gain an advantage in overcoming the protection of \emph{every} other utterance produced by the same speaker.  
The use case and approach to assessment may hence need additional thought. 

VoicePrivacy 2022 results show that \textbf{objective assessment} is \textbf{no substitute for subjective assessment}. 
The correlation in results for each is low; WER results for some utterances are near to one, while subjective intelligibility and naturalness scores for the same utterances are near to zero.  
Objective and subjective metrics are nonetheless both useful, but they reflect performance in different use cases.
The work in~\cite{champion_thesis} suggests that the retraining of an ASR system using anonymised data can result in mispronunciations still being correctly transcribed. 
WER estimates are then biased since they can mask mispronunciation errors.  
In the future we may have to consider distinct evaluation tracks or rankings which reflect anonymisation performance where the downstream task involves either automatic treatment of anonymised speech or consumption by human listeners (notwithstanding that some use cases may involve both).
In case of the latter, it may be necessary to conduct human listening transcription tests so that mispronunciations are properly taken into account.

\textbf{Challenge complexity}, including that of the baseline systems and evaluation framework, may account for why the challenge did not attract greater participation. 
%failed to attract the interest of the voice conversion community.
%, and may also account partly for why privacy research remains somewhat of a niche field; the cost of entry is too high. 
Ideas to simplify the baselines and evaluation framework involve the adoption of purely Python based solutions without reliance upon the integration of Kaldi components, and the clearer separation of baseline anonymisation components from those related to evaluation~\cite{meyer2023voicepat, franzreb23_spsc}.
With hope, these changes might attract participation from the voice conversion community.

%While processing capabilities will continue to expand in the future, the majority of current anonymisation solutions are highly demanding in terms of computation and memory and are therefore ill-suited to implementations for embedded or handheld devices, a realistic target use case for the publishing or sharing of personal, anonymised media.
%This calls for some reflection upon whether or not VoicePrivacy should somehow incentivise the development of lighter-weight anonymisation solutions alongside heavier-weight counterparts. 

%No line of work has yet considered the problem of \textbf{computational complexity}. While the field of speaker anonymisation might be too recent for such investigations to make sense, we point out that all state-of-the-art approaches rely on large neural network models that would be too cumbersome to run on embedded hardware found in smart devices.

%- a bigger problem than we realise - where is the VC community?  It's a free eval for them.s
%we simply must address the overall challenge complexity as well 
%\textbf{no more kaldi...?  purely python?  separate anon and eval packages?} % i don't think we'll have room for this either...

Last, we have foreseen for some time the development of a parallel \textbf{VoicePrivacy attacker challenge}.  
Appropriate strength testing is essential if we are to have any confidence in anonymisation safeguards.
Such an attacker challenge necessitates the collection and re-distribution of anonymised data, which would require the consent of VoicePrivacy challenge participants, in addition to the design of new evaluation protocols and, potentially, also metrics.  
A potential attacker challenge will likely follow the third edition of the traditional VoicePrivacy defender (anonymisation) challenge.  

%\vspace{10cm}  % please do NOT reduce this - we need space here to provide some direction for the future

\section{Conclusions}

VoicePrivacy 2022 attracted broad interest from the research community.
Eleven valid submissions were received, all of which
improved upon the challenge baselines to some degree, be it upon the primary privacy and utility measures, and/or upon other secondary measures such as voice distinctiveness or pitch correlation, etc. 
%in some specific aspect of the task of speaker anonymisation.
Given the primary motivation of fostering progress in voice anonymisation, VoicePrivacy 2022 was hence largely successful.
We must nonetheless acknowledge that we remain far from achieving real, effective voice anonymisation and that, with mounting privacy regulation, the community must redouble its efforts in the future. 

It is clear from the findings and our experience of VoicePrivacy 2022 that the organisation of challenges in voice anonymisation is itself \emph{challenging}. 
The use cases for anonymisation parallel those for speech technology more generally (e.g.\ the defined \emph{downstream} tasks).
Consequently, it is difficult to design a single challenge to foster progress in voice anonymisation in a manner that is suited to them all.
We must also acknowledge that voice anonymisation is but only one approach to preserve privacy in the use of speech technology and that, for some applications other, perhaps complementary techniques also warrant exploration. 
%For some use cases, speaker-level anonymisation is critical.  For others, utterance-level anonymisation is entirely sufficient. 

With alternative approaches being extremely diverse (e.g.\ encryption, federated learning, differential privacy etc.), VoicePrivacy will, at least for the time being, remain focused upon voice anonymisation.
In the penultimate section above, we outline our ideas for future challenge editions.
Our priorities will be to develop the challenge in a way which embraces more practical, commercial or societal use cases but also to simplify the challenge where possible, from the protocols, to the metrics, evaluation conditions and baseline systems. By lowering the cost of entry into what is still an emerging and demanding field of research, we hope to encourage broader participation. 
%Key to this goal will be 
%, while seeking convergence to real-world applications of speaker anonymisation.  
%We hope that this strategy will also help to attract broader participation. 
Given the increase in privacy regulation worldwide, we expect interest to grow rapidly in the coming years and with future challenge editions.

% \section*{Acknowledgement}

% JST-ANR VoicePersonae
% \vspace{3cm} % please do not reduce - we need the space for adding project acknowledgements later

\ifCLASSOPTIONcaptionsoff
  \newpage
\fi



%\begin{thebibliography}{1}


\bibliographystyle{IEEEtran}
\bibliography{IEEEabrv,Bibliography}

\vfill


\end{document}